\documentclass[twocolumn,pra]{revtex4}
\usepackage{color}
\usepackage{amsthm}
\usepackage{amsmath}
\usepackage{amssymb}
\usepackage{type1cm}
\usepackage[dvipdfmx]{graphicx}
\usepackage{txfonts}
\usepackage[english]{babel}
\usepackage{algorithm}
\usepackage{algorithmic}
\usepackage{amsfonts}

\newcommand{\ket}[1]{\left| #1 \right\rangle}
\newcommand{\bra}[1]{\left\langle #1\right |}

\begin{document}

\title{Asymmetric quantum multicast network coding: asymmetric optimal cloning over quantum networks}

\author{Yuichi Hirota}
\author{Masaki Owari}

\affiliation{$^{1}$ Department of Computer Science, Faculty of Informatics, Shizuoka University \\
3-5-1 Johoku, Naka-ku, Hamamatsu 432-8011, JAPAN}

\begin{abstract}
In this study, we consider a quantum version of multicast network coding as a multicast protocol for sending universal quantum clones (UQCs) from a source node  to the target nodes on a quantum network. 
By extending Owari et al.'s previous results for symmetric UQCs, we derive a protocol for multicasting  $1\rightarrow 2$ ($1\rightarrow 3$) {\it asymmetric} UQCs of a $q^r$-dimensional state to two (three) target nodes.
Our protocol works under the condition that each edge on a quantum network represented by an undirected graph $G$ transmits a $q$-dimensional state. There exists a classical solvable linear multicast network code with a source rate of $r$ on a classical network $G'$, where $G$ is an undirected underlying graph of an acyclic directed graph $G'$. We also assume free classical communication over a quantum network.
\end{abstract}

\maketitle

\section{Introduction}
The throughput of a network can be improved by applying non-trivial operations to the bitstream at intermediate nodes when there is a bottleneck on a network \cite{HL08,Y08}. This protocol can be referred to as  network coding. The network coding research was started in classical information theory \cite{ACLY00}. 
The network coding for a quantum network is called ``{\it quantum network coding}''  \cite{HINRY07}.
There has been a considerable amount of research on quantum network coding, which can improve the throughput of a quantum network in various situations \cite{Hayashi07,Shi06,Kobayashi09,Kobayashi10,Leung10,Kobayashi11,OKM13,KOM14,KOM15,EKB16}. 
Recently, it has been presented that quantum network coding can improve the security of a quantum network \cite{OKH17a,OKH17b,SH18a,SH18b}. 
Further, it is useful for quantum repeater networks \cite{SINV16,MSNV18} as well as for distributed quantum computation\cite{AM16}. Although many studies have considered network coding on noisy classical networks, almost all the studies of quantum network coding consider noise-free quantum networks. This is because 	quantum network coding is regarded as a protocol implemented on a layer after the errors have already corrected. 
Hence, in this study, we consider noise-free quantum networks.

\begin{figure}[h]
 \begin{center}
  \includegraphics[width=\linewidth]{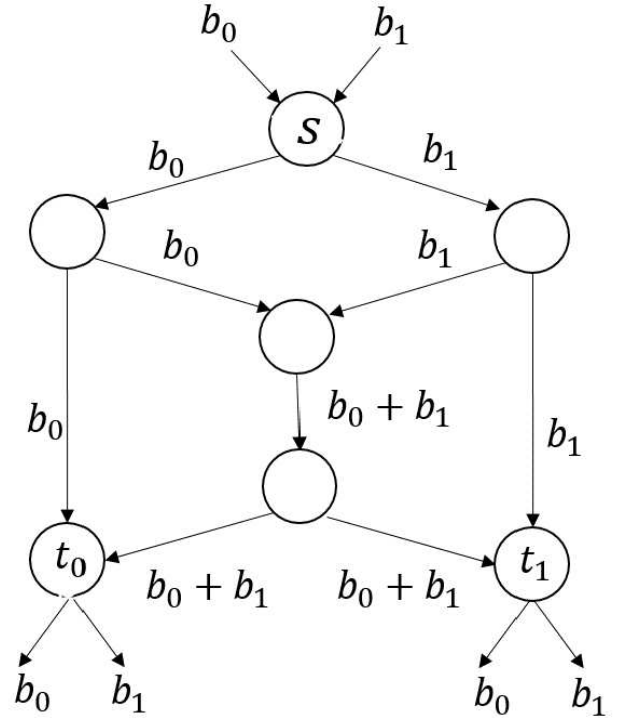}
 \end{center}
 \caption{Multicast  classical network coding on the butterfly network. A single source node $s$ sends messages $b_0$ and $b_1$ on $\mathbb{F}_p:=\mathbb{Z}/\mathbb{Z}_p$ to both target nodes $t_1$ and $t_2$.}
 \label{butterfly}
\end{figure}

\begin{figure}[h]
 \begin{center}
  \includegraphics[width=\linewidth]{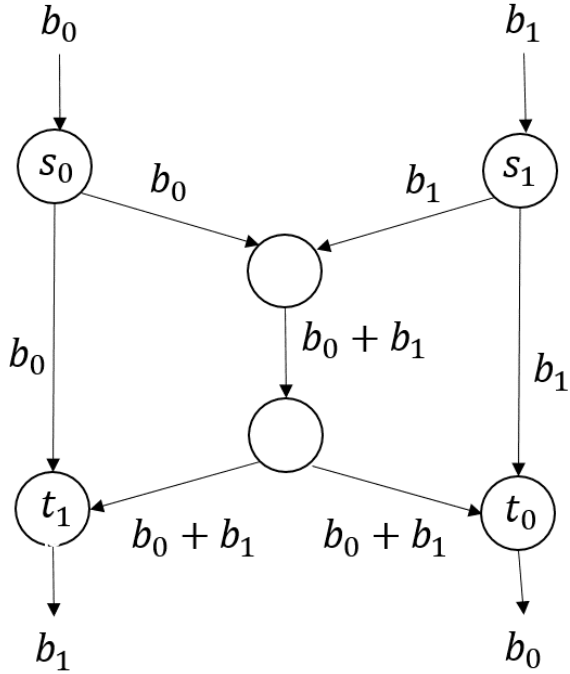}
 \end{center}
 \caption{Multiple-unicast classical network coding on the butterfly network. A source node $s_0$ sends message $b_0\in \mathbb{F}_p$ to target node $t_0$, and  source node $s_1$ sends message $b_1\in \mathbb{F}_p$ to target node $t_1$.}
 \label{butterfly2}
\end{figure}

In classical network coding, majority of the studies have focused on multicast communication, where a single source node transmits the same information to multiple target nodes on a network\cite{HL08,Y08}. Figure \ref{butterfly} shows the network coding for a the butterfly network.
This is one of the simplest examples of classical multicast network coding.
Another type of network coding is called multiple-unicast network coding.
Here, there are $k$ pairs of source and target nodes $(s_0,t_0 ),...,(s_{k-1},t_{k-1})$ on the network, and 
each source node $s_i$ independently  transmits a message to the corresponding target node $t_i$ for all $i$\cite{TRLKM06}. The modified version of the butterfly network in  Figure \ref{butterfly2} is one of the simplest examples of classical multiple-unicast network coding.

Most of the research on quantum network coding considered multiple-unicast communication, i.e., multiple-unicast quantum network coding. This differs from classical network coding because each source node transmits a quantum state (instead of a classical message) to the corresponding 
target node\cite{Hayashi07,Kobayashi09,Leung10,Kobayashi11,EKB16,OKH17a,OKH17b,SH18a,SH18b,SINV16,MSNV18}. 
The most important results are those of Kobayashi et al..
If classical information (or measurement results) can be freely sent among the nodes on a quantum network, Kobayashi et al. gave a canonical procedure for constructing a  quantum multiple-unicast network code from a classical multiple-unicast network code. The quantum network for the quantum code  and  the classical network for the classical code must represented by the same graph \cite{Kobayashi09,Kobayashi11}.

Unlike quantum multiple-unicast network coding, there has been less research on quantum network coding  focusing on multicast communication\cite{Shi06,Kobayashi10,OKM13,KOM14,KOM15}. 
This is because in quantum information theory, no-cloning theory prohibits perfect multicast communication \cite{WZ82}, and, thus,  it is not straightforward  to construct a multicast quantum network coding protocol as an extension of a  classical multicast network coding protocol. 

Shi et al.'s paper is the first to treat quantum multicast network coding \cite{Shi06}. They consider the problem of distributing $N$-identical copies of a state $\ket{\psi}$ from a single source node to $N$ target nodes. 
Since the number of copies of $\ket{\psi}$ is equal to the number of nodes, $\ket{\psi}$ can be distributed without cloning the quantum states. Shi et al. showed that coding on intermediate nodes can increase the throughput of the quantum network.  The second work treating this topic is Kobayashi et al.'s paper \cite{Kobayashi10}. In this paper, a single copy of  a state $\ket{\psi}=\sum _{i=1}^d \alpha_i \ket{i}$ is given on the source node and the aim is to share a Greenberger-Horne-Zeilinger (GHZ-)-type state $\sum _{i=1}^d \alpha_i \ket{i}_1 \otimes \cdots \otimes \ket{i}_N$ among target nodes, where the $i$th local system is on the $i$th target node. 
From this GHZ-type state between the target nodes, the input state $\ket{\Psi}$ can be reconstructed at any target node by local operations and classical communication (LOCC). 
Based on classical multicast network coding, Kobayashi et al. developed a quantum protocol to achieve the above task under the assumption of free classical communication among nodes on the quantum network. 

Although Shi et al.'s protocol and Kobayashi et al.'s protocol can be considered as generalizations of classical multicast network coding to quantum networks , rigorously speaking, the goal of their protocols is not exactly to achieve a multicast of a quantum state. Since an optimal multicast quantum channel is nothing but an optimal cloning\cite{BH96,SIG05,FWJYSZM14}, a protocol to share an optimal clone of an input state among target nodes of a quantum network can be considered as one of the most natural quantum extensions of a multicast classical network coding protocol. Based on this idea, Owari et al.\ constructed a protocol to share a symmetric optimal universal clone of an input state  on the target nodes under the conditions that classical information can be sent freely  among nodes on a quantum network and that a small amount of entanglement is shared on target nodes at the beginning of the protocol\cite{OKM13,KOM14,KOM15}. 

In this paper, we focus on extending Owari et al.'s results to asymmetric optimal  universal cloning\cite{NG98, C98, C00,IACFFG05}, which is a generalization of  symmetric optimal universal cloning.  Thus, we construct a protocol to efficiently multicast an asymmetric optimal clone of a $q^r$-dimensional input quantum state from one source node to two (three) target nodes, {\it where} $q$ {\it is assumed to be a prime power}.
In this protocol, the following five conditions are assumed:
\begin{itemize}
\item The noise free quantum network can be described by an undirected graph $G$ with one source node and two(three) target nodes.

\item Each quantum channel on the quantum network can transmit one $q$-dimensional quantum system in a single session.

\item There exists a classical solvable linear multicast network code with source rate $r$ for a noise-free classical network described by an acyclic directed graph $G'$, where $G$ is an undirected underlying graph of $G'$.

\item Measurement results (or classical information) can be sent freely from one node to another node on the quantum network.

\item  A small amount of entanglement which does not scale with $q$, is shared among the target nodes. The amount of entanglement is at most $2$\,ebit for  two target nodes, and  at most $\left( 2+4\log_2 3 \right )$\,ebit for the case of tree target nodes.
\end{itemize}
Using the max-flow and min-cut theorem of multicast network coding \cite{HL08,Y08}, for sufficiently large $q$, the assumption for the  existence of a classical network code on $G'$  can be replaced by  the condition that the minimum-cut between the source node $s$ and a target node $t_i$ is no less than $r$ for all $i$. 

\begin{figure}[h]
 \begin{center}
  \includegraphics[width=90mm]{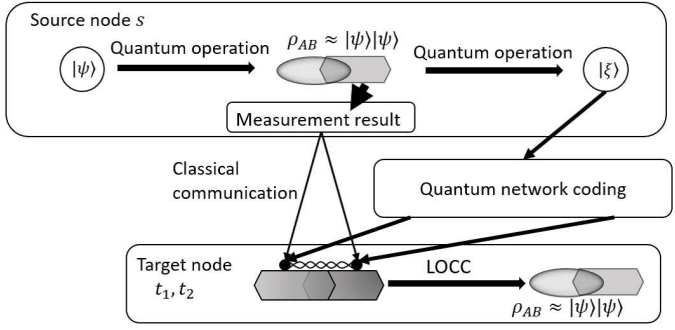}
 \end{center}
 \caption{Schematic diagram of a protocol for multicasting asymmetric optimal clones from one source node to two target nodes.
The asymmetric optimal cloning protocol for the input state $\ket{\psi}$ is implemented at the source node.ng
The resulti state is compressed into a $d$-dimensional system, and transmitted to the two target nodes using Kobayashi et al.'s quantum multicast network coding protocol. Finally, the asymmetric clones of the input state are reconstructed by LOCC on target nodes with the help of a small amount of entanglement.}
 \label{asym_multi_sketch}
\end{figure}

An outline of our protocol is as follows:
\begin{itemize}
\item We creat two (three) asymmetric optimal clones of an input state with an ancilla system at a source node. 

\item We measure the ancilla system, and send the measurement outcomes to the target nodes.

\item We compress the whole $d^2$ ($d^3$)-dimensional system into a $d$-dimensional system. 

\item  We transmit the resulting state to two(three) target nodes using Kobayashi et al.'s multicast quantum network coding\cite{Kobayashi10}. As a result, a GHZ-type state is shared among target nodes.

\item We reconstruct the asymmetric clones of the input state from the GHZ-type state
using LOCC with a small amount of entanglement among the target nodes and the measurement outcomes sent from the source node.
\end{itemize}
Using the above protocol, we can multicast  asymmetric optimal clones from one source node to two(three) target nodes (Figure~\ref{asym_multi_sketch}).

The rest of paper is organized  as follows:
We explain the asymmetric cloning and Kobayashi et al.'s quantum multicast network coding protocol in Section~\ref{sec:ii}.
We present a procol for multicasting $1\rightarrow 2$ asymmetric optimal clones in Section~\ref{sec:iii}. We also present a procol for multicasting $1\rightarrow 3$ asymmetric optimal clones in Section~\ref{sec:iv}.
Finally, we gives an conclusion in Section~\ref{sec:v}.

\section{Preliminaries}\label{sec:ii}
Optimal asymmetric universal quantum cloning, classical linear multicast network coding and Kobayashi et al.'s multicast quantum network coding protocol are all important  in our protocol. In this section, we explain optimal asymmetric  universal quantum cloning  in Section~\ref{u_cloning}. Then, classical linear multicast network coding and the Kobayashi et al.'s multicast quantum network coding protocol are presented in Sections~\ref{subsec classical} and~\ref{q_net_coding}, respectively.

\subsection{Optimal asymmetric quantum universal cloning machine \label{u_cloning}}

No-cloning theorem states that quantum mechanics prohibits a quantum operation that makes perfect copies of an unknown quantum state\cite{WZ82}.
In other words, a perfect multicast of  an unknown quantum state is impossible.
On the other hand, quantum mechanics does not completely prohibit the approximate cloning of a quantum state. 
Hence, many studies have focused on quantum protocols to make an approximate copy of unknown states (so called quantum cloning machines) \cite{BH96,SIG05,FWJYSZM14,C00,IACFFG05}.

A quantum cloning machine (QCM) that produces  $N$ approximate clones based on $M$ copies of a given quantum state $\ket{\psi}\in \mathcal{H}$ is 
a quantum channel (or a completely positive and trace preserving map) $\varepsilon$ from $\mathfrak{B} \left( \mathcal{H}^{\otimes M}\right )$ to $\mathfrak{B} \left( \mathcal{H}^{\otimes N}\right)$, where $\mathfrak{B} \left( \mathcal{H} \right)$ is a space of all linear operators on the Hilbert space $\mathcal{H}$. 
Suppose $\rho_i$ is a reduced density matrix of the output state on the $i$th subsystem: $\rho_i = \operatorname{Tr}_{\neg i}\varepsilon \left( \left( \ket{\psi}\bra{\psi} \right)^{\otimes M} \right)$, where $\operatorname{Tr}_{\neg i}$ is a partial trace of all subsystems except the $i$th subsystem. 
Since the purpose of a QCM is to make $\rho_i$ as closed as the input state $\ket{\psi}\bra{\psi}$, the performance of a QCM can be described by the output fidelity $F_{i}$ between $\rho_i$ and $\ket{\psi}\bra{\psi}$:
\begin{eqnarray}
F_i=\langle \psi| \rho_i | \psi \rangle, \ \ (i=1,...,M).
\end{eqnarray}

A QCM is called universal, if $F_{i}$ does not depend on the input state $\ket{\psi}$. Further, a universal QCM (UQCM) is called symmetric, if the all clones are the same: $\rho_i = \rho_j$ for all $i$ and $j$. A UQCM that is not symmetric is asymmetric. An asymmetric UQCM whose output fidelities $F_i$ are optimal is called an optimal asymmetric UQCM. Since the output states of an asymmetric UQCM satisfy $\rho_i \neq \rho_j$, the output fidelities $F_i$ also depend on $i$.  Hence, an optimal asymmetric UQCM in general depends on parameters that represent a bias among the output fidelities $\{ F_i \}_{i=1}^N$.

Here, we give an optimal asymmetric UQCM with $M=1$ and $N=2$ (we call this protocol a $1 \rightarrow 2$ optimal asymmetric UQCM).
This protocol uses three systems $A$, $B$, and $M$ whose Hilbert spaces are $\mathcal{H}_A$,  $\mathcal{H}_B$, and $\mathcal{H}_M$, respectively. Here $\mathcal{H}_A$ works as an input system and a first output system, $\mathcal{H}_B$ is a second output system, and $\mathcal{H}_M$ is an ancilla system. The dimensions of all three systems are the same, and we denote this dimension as $d$; that is, $\dim \mathcal{H}_A=\dim \mathcal{H}_B=\dim \mathcal{H}_M=:d$.
Then, for an input state $\ket{\psi}$ on system $A$, a $1\rightarrow 2$ optimal asymmetric UQCM is given by an isometry $U_{1\rightarrow 2}^{(a,b)}$ from $\mathcal{H}_A$ to $\mathcal{H}_A \otimes \mathcal{H}_B \otimes \mathcal{H}_M$ satisfying \cite{C00}:
\begin{eqnarray}
 U_{1\rightarrow 2}^{(a,b)} |\psi \rangle_A  = a|\psi \rangle_A|\Phi_{d}^{+}\rangle_{BM} + b| \psi \rangle_B|\Phi_{d}^{+}\rangle_{AM} \label{1+1}.
\end{eqnarray}
where $\ket{+}$ is defined by $\ket{+}: = \frac{1}{\sqrt{d}}\sum_{k=0}^{d-1}|k \rangle_M$, $\ket{\Phi_d^+}$ is a standard $d$-dimensional maximally entangled state:
\begin{eqnarray}
|\Phi_{d}^{+}\rangle : = \frac{1}{\sqrt{d}}\sum_{k=0}^{d-1} |k \rangle|k \rangle,
\end{eqnarray}
and $a$ and $b$ are real parameters satisfying: 
\begin{equation}
a^2+b^2+\frac{2ab}{d}=1.
\end{equation}
Using $U_{1\rightarrow 2}$ defined above, the optimal asymmetric UQCM $\varepsilon_{1\rightarrow 2}^{(a,b)}$ is 
\begin{equation}\label{eq asymmetric uqcm}
\varepsilon_{1\rightarrow 2}^{(a,b)} \left(\ket{\psi}\bra{\psi}\right) := \operatorname{Tr}_M \left( U_{1\rightarrow 2}\ket{\psi}\bra{\psi}_A U_{1\rightarrow 2}^\dagger \right). 
\end{equation}
The fidelity of the reduced density matrices, which have been proved to be optimum \cite{C00}, are given by 
\begin{align}
F_{A}& := \bra{\psi}\mathrm{Tr} _{B} \left( \varepsilon_{1\rightarrow 2}^{(a,b)} \left(\ket{\psi}\bra{\psi}\right)\right) \ket{\psi}=1-b^2\frac{d-1}{d}, \nonumber \\
F_{B}& := \bra{\psi}\mathrm{Tr} _{A} \left( \varepsilon_{1\rightarrow 2}^{(a,b)} \left(\ket{\psi}\bra{\psi}\right)\right) \ket{\psi} =1-a^2\frac{d-1}{d}.
\end{align}

Next, we give an optimal asymmetric UQCM with $M=1$ and $N=3$ (we call this protocol the $1 \rightarrow 3$ optimal asymmetric UQCM).
This protocol use five systems $A$, $B$, $C$, $R$, and $S$ whose Hilbert spaces are $\mathcal{H}_A$,  $\mathcal{H}_B$,  $\mathcal{H}_C$,  $\mathcal{H}_R$,  and $\mathcal{H}_S$, respectively. Here, $\mathcal{H}_A$ is an input system that is also the first output system. $\mathcal{H}_B$  and $\mathcal{H}_C$  are the second and third output systems, respectively. $\mathcal{H}_R$  and $\mathcal{H}_S$ are ancilla systems. The dimensions of all systems are the same, which we denote as $d$. For an input state $\ket{\psi}$ on system $A$, $1 \rightarrow 3$ optimal asymmetric UQCM is given by an isometry $U_{ABCRS}$ 
from $\mathcal{H}_A$ to $\mathcal{H}_A \otimes \mathcal{H}_B \otimes\mathcal{H}_C \otimes\mathcal{H}_R \otimes\mathcal{H}_S$ satisfying the following equation:  
\begin{align}
& U_{1 \rightarrow 3}^{(\alpha, \beta, \gamma) } \ket{\psi} \nonumber \\
= & \sqrt{\frac{d}{2d+2}}[\alpha | \psi \rangle_A(|\Phi^{+} \rangle_{BR}|\Phi^{+} \rangle_{CS}+ |\Phi^{+} \rangle_{BS}|\Phi^{+} \rangle_{CR}) \nonumber \\
&\quad + \beta|\psi \rangle_B(|\Phi^{+}\rangle_{AR}|\Phi^{+}\rangle_{CS} + |\Phi^{+}\rangle_{AS}|\Phi^{+}\rangle_{CR})   \nonumber \\
& \quad + \gamma|\psi \rangle_C(|\Phi^{+} \rangle_{AR}|\Phi^{+} \rangle_{BS} + |\Phi^{+} \rangle_{AS}|\Phi^{+} \rangle_{BR})] \label{1+1+1},
\end{align}
where $\alpha, \beta, \gamma$ are non-negative real parameters satisfying the following constraint \cite{FWJYSZM14, IACFFG05}:
\begin{equation}
\alpha^{2} + \beta^{2} + \gamma^{2} + \frac{2}{d}(\alpha \beta + \beta \gamma + \alpha \gamma) = 1.
\end{equation} 
In terms of $U_{ABCRS}$, a $1\rightarrow3$ optimal asymmetric UQCM $\varepsilon_{1 \rightarrow 3}^{(\alpha, \beta, \gamma) }$ can be written as:
\begin{equation}\label{eq def asymmetric uqcm 13}
\varepsilon_{1 \rightarrow 3}^{(\alpha, \beta, \gamma) } \left(\ket{\psi}\bra{\psi}\right) := \mathrm{Tr}_{RS} \left( U_{1 \rightarrow 3}^{(\alpha, \beta, \gamma) }\ket{\psi}\bra{\psi}_A U_{1 \rightarrow 3}^{(\alpha, \beta, \gamma) \ \dagger } \right). 
\end{equation}
The fidelities between an input state and each reduced density matrix, which were proved to be optimum \cite{FWJYSZM14, IACFFG05}, is given as follows: 
\begin{align}
F_A & = 1-\frac{d-1}{d}\left( \beta ^2 + \gamma ^2 + \frac{2 \beta \gamma}{d+1} \right), \nonumber \\
F_B & = 1-\frac{d-1}{d}\left( \alpha ^2 + \gamma ^2 + \frac{2 \alpha \gamma}{d+1} \right), \nonumber \\
F_A & = 1-\frac{d-1}{d}\left( \alpha ^2 + \beta ^2 + \frac{2 \alpha \beta}{d+1} \right).
\end{align}

\subsection{Classical multicast network coding}\label{subsec classical}
Since our protocol uses Kobayashi et al.'s protocol as a subroutine and since Kobayashi
et al.'s protocol is based on a classical linear multicast network code, we introduce classical linear multicast network coding  in this section. A detail description of classical multicast network coding can be found in standard text books of network coding like \cite{HL08,Y08}.

A classical network is represented by a directed graph $G'=(V, E')$, 
where a vertex $v \in V$ represents a node of the network and an edge $e \in E'$ represents a noiseless classical channel. In this paper, we assume that $G'$ is {\it acyclic}. 
There exist a source node $s \in V$, and $N$ target nodes $t_1, \dots, t_N \in V$ on the network. A node that is neither a source node nor a target node is called an intermediate node.
In a single session of a classical multicast network coding, an alphabet on the finite field $\mathbb{F}_q$ is sent from node $u$ to node $v$ if $(u,v) \in E'$, where the order of $\mathbb{F}_q$ is a prime power $q$. 
Since $G'$ is an acyclic directed graph, 
a natural partial ordering can be defined on $E'$;
that is, when $(u, v), (v, w) \in E'$, we define $(u, v) \prec (v,w)$. 
 The order of transmissions of classical information can be determined by this partial ordering. That is, an edge $e \in E'$ transmits an alphabet after all edges $e'\in E'$ satisfying $e' \prec e$ have transmitted alphabets. 
We assume that there is no incoming edge to the source node $s$, and that
there is no outgoing edge from any target node. 
Hence, all edges whose tail node is the source node $s$ are a local minimum, 
and all edges whose head node is a target node are a local maximum under the partial ordering. We further assume that all edges whose tail node is not the source node $s$ are not a local minimum and that all edges whose head node is not a target node are not a local maximum.

A classical linear multicast network code over $\mathbb{F}_q$ on $G'$ consists of a set of linear maps $\{ f_e \}_{e \in E'}$.
At the beginning of a session, an input message $\vec{x}:=\left( x_1, \dots, x_r \right) \in \mathbb{F}_q^r$ is chosen on the source node $s$, where $r$ is the {\it source rate} of the classical multicast network code.
Suppose $e$ is an outgoing edge of $v$. At the first step of the network coding, 
an alphabet $y_e$ transmitted through the edge $e$ is chosen as a linear combination of $x_1, \dots, x_r$. In other words, in terms of a linear function $f_e: \mathbb{F}_q^r \rightarrow \mathbb{F}_q$, $y_e$ can be written as 
\begin{eqnarray}
y_e := f_e(\vec{x}) = f_e(x_1,\cdots x_r).
\end{eqnarray}
After calculating  $y_e$, $y_e$ is transmitted through $e$. 
After all edges outgoing from the source node $s$ transmitted an alphabet,  all intermediate nodes transmit alphabet in the order determined by the partial ordering as follows:
Suppose an intermediate node $v$ on the network has $m$ incoming edges and $e$ is an outgoing edge from $v$. 
After all transmissions of $m$ incoming edges to $v$ have finished, 
the node $v$ has $m$-alphabets $y_j \in \mathbb{F}_{q}$ ($j = 1, \dots ,m$), where $y_j$ is an alphabet sent through the $j$th incoming edge.
Then, an alphabet $y_e$ transmitted through the edge $e$ is chosen as a linear combination of $y_1, \dots, y_m$. In other words, there exists a linear function $f_e: \mathbb{F}_q^m \rightarrow \mathbb{F}_d$ such that 
\begin{eqnarray}
y_e := f_e(y_1,\cdots y_m) \label{nc}.
\end{eqnarray}
After the calculation, $y_e$ is transmitted through $e$. 

Suppose a target node $t_i$ has $m_i$ incoming edges. Then, after all edges have transmitted an alphabet, the target node $t_i$ has $m_i$-alphabets $y_j^{(i)} \in \mathbb{F}_{q}$ ($j = 1, \cdots ,m$), where $y_j^{(i)}$ is an alphabet sent through the $j$th incoming edge to $t_i$. A classical linear multicast network code $\{ f_e \}_{e \in E'}$ is called {\it solvable} if there exists a set of decoding operations $\{ g_i \}_{i=1}^N$ such that $g_i: \mathbb{F}_q^{m_i} \rightarrow \mathbb{F}_q^r$ satisfies the following equation for all $i$:
\begin{equation}
\vec{x} = g_i\left( y_1^{(i)}, \cdots, y_{m_i}^{(i)} \right),
\end{equation}
where $\vec{x}\in \mathbb{F}^r_q$ is the input message. If a classical linear multicast network code is solvable, any decoding operation $g_i$ can be chosen as a linear map. 

There is a necessary and sufficient condition for the existence of a classical linear multicast network code \cite{HL08,Y08}.   Suppose $C_i$ is the size of the minimum cut between $s$ and $t_i$. 
Then, there exists a classical linear multicast code with source rate $r$ on $G'$ over a sufficiently large field $\mathbb{F}_q$, if and only if $C_i \ge r$ for all $i$.

\subsection{Quantum multicast  network coding\label{q_net_coding}}
In this section, we review Kobayashi et al.'s protocol \cite{Kobayashi10}. 
First, we give  a problem setting for multicast quantum network coding 
that is common between our protocol and Kobayashi et al.'s protocol. 
A quantum network is described by an {\it undirected} graph $G=\left( V, E \right)$, where $V$ represents a set of nodes and $E$ represents a set of quantum channels. There exist a source node $s \in V$, and $N$ target nodes $t_1, \dots, t_N \in V$ on the network. 
In a single session, any quantum channel $(u,v) \in E$ can send a $q$-dimensional quantum system $\mathcal{H}_e$ just once either from $u$ to $v$, or from $v$ to $u$,
where $q$ is assumed to be a prime power.
Further, any quantum operations can be implemented on any node $v \in V$, and measurement outcomes (or classical information) can be  freely sent among nodes. 
At the beginning of a session, a single copy of input state $\ket{\psi}$ is given on the source node $s$. 
Here, the reason a quantum channel is represented by an undirected edge is that  the direction of a quantum channel can be effectively reversed by quantum teleportation under the condition of free classical communication\cite{Leung10}.

The purpose of both protocols is to multicast an input state $\ket{\psi}$ from the source node to all target nodes in a single session. 
Here, we should note that the meaning of ``multicast'' in Kobayashi et al.'s protocol is different from that in our protocol. 
As we have explained in the introduction, the purpose of our protocol is to construct optimal asymmetric universal clones among target nodes for a given $q^r$-dimensional input state $\ket{\psi}=\sum _{j=0}^{q^r-1} \alpha_j \ket{j} \in \mathcal{H}_s$ on a source node, where $\mathcal{H}_s$ is a $q^r$-dimensional input space.  In other words, we consider multicast quantum network coding with source rate $r$.  On the other hand, the purpose of Kobayashi et al.'s protocol is to construct a GHZ-type state $\sum _{j=0}^{q^r-1} \alpha_j \ket{j}_1 \otimes \cdots \otimes \ket{j}_N$ among target nodes, where the $i$th local system is on the $i$th target node. 

Both Kobayashi et al.'s protocol and our protocol are constructed under the assumption that there exists a solvable classical linear multicast network code $\{f_e \}_{e \in E'}$ with source rate $r$ on an acyclic directed graph $G'=(V, E')$ over a finite field $\mathbb{F}_q$, where $G$ is an undirected underlying graph of $G'$. In other words, $G$ can be derived by replacing all directed edges on $G'$ by undirected edges.  Using this replacement, a directed edge $e' \in E'$ is naturally mapped to an undirected $e \in E$, and this map is a bijection. Hence, in the following part of this section, we will not distinguish $e'$ from $e$, and write $e'$ as $e$.  

Kobayashi et al.'s protocol imitates a classical linear multicast network code $\{ f_e \}_{e \in E'}$ and corresponding decoding operations $\{ g_i \}_{i=1}^N$ by unitary operators. Each linear map $f_e$ is imitated by a unitary operator $U_e$, and each recovery operator $g_i$ is imitated by a unitary operator $V_i$, where $U_e$ and $V_i$ are defined as follows: 
Since $\dim \mathcal{H}_s =q^r$, due to the  bijection between $\{0, 1 \dots q^r-1 \}$ and $\mathbb{F}_q^r$, an input state $\ket{\psi} \in \mathcal{H}_s$ can be written as $\ket{\psi}=\sum _{\vec{x} \in \mathbb{F}_q^r} \alpha _{\vec{x}} \ket{\vec{x}}$.
For an outgoing edge $e$  from the source node $s$, a unitary operator $U_e$ on $\mathcal{H}_s \otimes \mathcal{H}_e$ is defined  by means of $f_e: \mathbb{F}_q^r \rightarrow \mathbb{F}_q$ as 
\begin{equation}
U_e:= \sum _{\vec{x}\in \mathbb{F}_q^r, y \in \mathbb{F}_q}
\ket{\vec{x}}\bra{\vec{x}}_s \otimes \ket{y+f_e\left (\vec{x} \right) }\bra{y}_e,
\end{equation}
where $\mathcal{H}_e$ is a Hilbert space transmitted through $e$.
Suppose $In(e)$ is a set of all incoming edges of $v$, where $v$ is a tail node of $e$, and suppose $\mathcal{H}_{In(e)}:= \bigotimes _{e' \in In (e)}  \mathcal{H}_{e'} $. Then, for an outgoing edge $e$ from an intermediate node $v$, a unitary operator $U_e$ on $\mathcal{H}_{In(e)} \otimes \mathcal{H}_e$ is defined  by means of $f_e: \mathbb{F}_q^{|In(e)|} \rightarrow \mathbb{F}_q$ as 
\begin{equation}
U_e:= \sum _{\vec{y} \in \mathbb{F}_q^{|In(e)|}, y_e \in \mathbb{F}_q}
\ket{\vec{y}}\bra{\vec{y}}_{In(e)} \otimes \ket{y_e+f_e\left (\vec{y} \right ) }\bra{y_e}_e.
\end{equation}
Suppose $\mathcal{V}_i$ is a $q^r$-dimensional output Hilbert space on a target node $t_i$. A unitary operator $V_e$ on $\mathcal{H}_{In(t_i)} \otimes \mathcal{V}_i$ is defined  by means of the decoding operation $g_i: \mathbb{F}_q^{|In(t_i)|} \rightarrow \mathbb{F}_q^r$ as 
\begin{equation}
V_i:= \sum _{\vec{y} \in \mathbb{F}_q^{|In(t_i)|}, \vec{x} \in \mathbb{F}_q^r}
\ket{\vec{y}}\bra{\vec{y}}_{In(t_i)} \otimes \ket{\vec{x}+g_i\left (\vec{y} \right ) }\bra{\vec{x}}_i.
\end{equation}

Kobayashi et al.'s quantum multicast network coding protocol is shown as protocol 1.
\begin{Protocol}[H]
\caption{Kobayashi et al.'s quantum multicast network coding protocol}         
\begin{algorithmic}
\STATE \textbf{ Step 1: Initialization}

The source node $s$ prepares an initial state $\ket{\psi}$ on $\mathcal{H}_s$.
Each node $v \in V$ prepares $\ket{0}$  on $\mathcal{H}_e$ for an edge $e \in E$ whose tail node is $v$. 
For all $i$ satisfying $1 \le i \le m$, a target node $t_i$ prepares $\ket{0}$ on 
$\mathcal{V}_i$.

\STATE \textbf{Step 2: Transmission}

First, for all edges $e \in E'$ whose tail node is the source node, the source node operates the unitary operator $U_e$ on  $\mathcal{H}_s \otimes \mathcal{H}_e$ and sends $\mathcal{H}_e$ to the head node of $e$. 
Second, all intermediate nodes behave in the order defined by the natural partial ordering on $E'$ as follows: After an intermediate node $v$ has received Hilbert spaces from all edges whose head node is $v$,
for all edges $e \in E'$ whose tail node is $v$, node $v$ operates the unitary operator $U_e$ on  $\mathcal{H}_{In(e)} \otimes \mathcal{H}_e$ and sends $\mathcal{H}_e$ to the head node of $e$. 
Finally, after all edges have transmitted Hilbert spaces, 
for all $i$ satisfying $1\le i \le m$, target node $t_i$ operates the unitary operator $V_i$ on  $\mathcal{H}_{In(t_i)} \otimes \mathcal{V}_i$.

\STATE \textbf{Step3: Measurement on Fourier-basis}

The source node $s$ measures the Hilbert space $\mathcal{H}_s$ in the Fourier basis, and sends  the measurement outcome to all the terminal nodes $t_i$.
For all edges $e \in E'$, the head node of $e$ 
measures the Hilbert space ${\mathcal{H}}_{e}$ in the Fourier basis,
and sends the measurement outcome to all terminal nodes $t_i$. 

\STATE \textbf{Step 4: Recovery}

All terminal nodes $t_i$ operate $Z(c_1)\otimes \cdots \otimes Z(c_r)$ on $\mathcal{V}_i$. Here, $\{ c_k \}_{k=1}^r$ is a natural number that can be determined from the measurement outcomes received in step 3, the classical linear multicast network code $\{ f_e \}_{e\in E'}$, and the decoding operators $\{ g_i \}_{i=1}^N$ \cite{Kobayashi10}.

\end{algorithmic}
\end{Protocol}
In step 3 of protocol 1, the Fourier basis of $\left\{ \ket{\tilde{z}} \right\} _{z \in \mathbb{F}_q}
 \subset {\mathcal{H}_e}$
of the computational basis $\left\{ \ket{x} \right\} _{x \in \mathbb{F}_q} \subset{\mathcal{H}}_{e}$
is defined 
as 
\[
\ket{\tilde{z}} :=\sum_{x \in \mathbb{F}_q} \omega^{\operatorname{Tr} xz }\ket{x},
\]
where $\omega:=\exp\left(-2\pi i / p \right)$.
Here, $\operatorname{Tr} z $ represents 
the element $\operatorname{Tr} M_z \in \mathbb{F}_p$, 
where $M_z$ is the matrix representation of the multiplication map $x \mapsto zx$. 
Here, we note that the finite field $\mathbb{F}_{q}$ can be identified with the vector space $\mathbb{F}_p^t$, where $t$ is the degree of the algebraic extension of $\mathbb{F}_{q}$.
For further details, see \cite[Section 8.1.2]{Haya2}.
We also define the generalized Pauli operators $Z(t)$ 
as  $Z(t):=\sum_{x \in \mathbb{F}_q}\omega^{\operatorname{Tr} xt}\ket{x}\bra{x}$.

\section{$1\rightarrow 2$ asymmetric UQC multicast protocol }\label{sec:iii}
	In this section, we present a new protocol that multicasts optimal asymmetric UQCs from the source node $s$ to two target nodes $t_1$ and $t_2$ on a quantum network.  We present the protocol in Section~\ref{protocol 1 2} and prove that the it creates optimal asymmetric UQCs in the subsection \ref{subsec proof 1 2}.

\subsection{$1\rightarrow 2$\ quantum multicast protocol}\label{protocol 1 2}
In this section, we present the protocol for multicasting $1\rightarrow 2$ optimal asymmetric UQCs of an input quantum state  from  the source node $s$ to two target nodes $t_1$ and $t_2$.  

As we have explained in Section~\ref{q_net_coding}, the problem settings for Kobayashi et al.'s protocol and our protocol are essentially the same, and the only their purposes are different. 
Here, we summarize the problem setting of our quantum multicast network coding:
A quantum network is described by an {\it undirected} graph $G=\left( V, E \right)$. There exist a source node $s \in V$, and $N$ target nodes $t_1, \dots, t_N \in V$ on the network. In this section, since we consider multicasting $1\rightarrow 2$ asymmetric UQCs, we set $N=2$.
In a single session, any quantum channel $(u,v) \in E$ can send a $q$-dimensional quantum system $\mathcal{H}_e$ just once, either from $u$ to $v$ or from $v$ to $u$,
where $q$ is assumed to be a prime power.
Further, any quantum operations can be implemented on any node $v \in V$, and measurement outcomes can be  freely sent among nodes. 
At the beginning of a session, a single copy of input state $\ket{\psi}$ is given on the source node $s$. 

Under these problem settings, the purpose of our protocol is to construct optimal asymmetric universal clones given by Eq.~(\ref{eq asymmetric uqcm}) between target nodes $t_1$ and $t_2$ for a given $d$-dimensional input state $\ket{\psi}=\sum _{j=0}^{d} \alpha_j \ket{j} \in \mathcal{H}_s$ on a source node, where $\mathcal{H}_s$ is a $d$-dimensional input space.
We assume $d=q^r$. In other words, we consider multicast quantum network coding with source rate $r$.
Here, note that since we assumed $q$ is a prime power, $d$ is also a prime power. 

For this purpose, we use two additional assumptions:
The first assumption is the same assumption that Kobayashi et al.\ used. That is, we assume that there exists a solvable classical linear multicast network code $\{f_e \}_{e \in E'}$ with source rate $r$ on an acyclic directed graph $G'=(V, E')$ over a finite field $\mathbb{F}_q$, where $G$ is an undirected underlying graph of $G'$.   
Hence, we can use Kobayashi et al.'s quantum multicast network coding protocol 
with source rate $r$ on this quantum network $G$.  We further assume that at most $2$\,ebits of entanglement resource are shared between target node $t_1$ and $t_2$. Hence, the amount of this entanglement resource is constant with respect to the dimension $d$
of the input state, and is negligible for large $d$ in comparison to $d$.   

Before we present the protocol, we define the unitary operators used in it. Pauli operators $X_d$ and $Z_d$ are defined as 
\begin{equation} 
X_d:=\sum _{k=0}^{d-1}\ket{k \oplus 1}\bra{k},
\quad Z_d:=\sum _{k=0}^{d-1}\omega^k \ket{k }\bra{k},
\end{equation}
where $\omega := e^{\frac{2\pi i}{d}}$.
In the following part of the paper, unitary operators defined on 
$\mathbb{C}^d \otimes \mathbb{C}^d$ and $\mathbb{C}^d \otimes \mathbb{C}^d \otimes \mathbb{C}^d$ are called bipartite and tripartite unitary operators, respectively.
$\Upsilon^{(r)}$ is defined as  a bipartite unitary operator satisfying 
\begin{align}\label{eq def upsilon 1}
\Upsilon^{(r)} \left( \cos \eta|jr \rangle + \sin \eta|rj \rangle \right) = |jr \rangle,
\quad  
\Upsilon^{(r)} |rr \rangle =  |rr \rangle, \nonumber \\
\Upsilon^{(r)} \left( \sin \eta|jr \rangle - \cos \eta |rj \rangle \right)  = |rj \rangle \qquad 
\end{align}
for all $j \in \{ 0, \dots, d-1 \}$ satisfying $j \neq r$, and 
\begin{equation}\label{1+1clone}
\Upsilon^{(r)} \ket{ij} = \ket{ij}
\end{equation} 
for all $i,j \in \{0, \dots, d-1 \}$ satisfying $i,j \neq r$, where $\eta$ is defined by 
\begin{equation}\label{eq def eta}
\cos \eta = \frac{a}{\sqrt{1-\frac{2ab}{d}}} \quad \text{and} \quad  \sin \eta = \frac{b}{\sqrt{1-\frac{2ab}{d}}}.
\end{equation}
The bipartite unitary operator $V^{(r)}$ is defined by 
\begin{equation}\label{eq def V r}
V^{(r)} := \sum_{j \neq r} |j \rangle \langle j| \otimes U_{r,j} + |r \rangle \langle r | \otimes I
\end{equation}
where the unitary operator $U_{r,j}$ is defined by 
\[U_{r,j}=I - |j \rangle \langle j| - |r-1 \rangle \langle r-1| + |j \rangle \langle r-1 | + |r-1 \rangle \langle j |.\]
The bipartite unitary operator $\Delta^{(r)} $ is  defined by 
\begin{equation}\label{eq def Delta r}
\Delta^{(r)} := |r \rangle \langle r | \otimes X_d^{-(r-1)} + \sum_{j \neq r}|j \rangle \langle j| \otimes I.
\end{equation}
The tripartite unitary operator $\Gamma^{(r)}$ is defined by
\begin{equation}\label{eq def Gamma r}
\Gamma^{(r)} := |r \rangle \langle r | \otimes \operatorname{swap} + \sum_{j \neq r}|j \rangle \langle j| \otimes I,
\end{equation}
 where $\operatorname{swap}$ is a unitary operator on $\mathbb{C}^d \otimes \mathbb{C}^2$ defined by  \[\operatorname{swap} := \sum_{i = 2}^{d-1} \sum_{j=0,1} |ij \rangle \langle ij| + \sum_{i,j=0,1} |ij \rangle \langle ji |.\]
The unitary operator $\Theta$ on $\mathbb{C}^2 \otimes \mathbb{C}^2$ is defined by
\begin{align}
\Theta \ket{jj} & = \ket{ jj} \quad (j = 0, 1) \nonumber  \\
\Theta \left (\cos \eta \ket{01} + \sin \eta  \ket{10} \right ) & =
 \ket{10} \nonumber \\
\Theta \left ( \sin \eta \ket{01} - \cos \eta  \ket{10} \right ) & =
\ket{01}   \label{step9}
\end{align}
The bipartite unitary operator $\Lambda ^{(r)}$ is defined by 
\begin{equation}\label{eq def lambda}
\Lambda^{(r)} := \sum_{j \neq r} |j \rangle \langle j| \otimes I + |r \rangle \langle r | \otimes X^r
\end{equation}

Before starting the protocol, we prepare three $d$-dimensional systems $A$, $B$ and $M$ at the source node $s$, 
$d$-dimensional systems $C$, $E$ and  $2$-dimensional systems $G$, $T_1$ at the target node $t_1$.
Similarly, we prepare $d$-dimensional systems $D$, $F$ and $2$-dimensional systems $H$, $T_2$ at $t_2$.
The entanglement resource $\cos \eta |0\rangle_E |1 \rangle_E + \sin \eta|1 \rangle_E |0 \rangle_F$ is shared between $E$ and $F$,
and the Bell state $\frac{1}{\sqrt{2}}(|00\rangle_{T_1 T_2} + |11\rangle_{T_1 T_2})$ is shared between $T_1$ and $T_2$.
Thus, the amount of entanglement resources is at most $2$\,ebits.

\begin{Protocol}[!h]
\caption{$1\rightarrow 2$\ quantum multicast network coding protocol}         
\begin{algorithmic}
\STATE \textbf{Step 1:} 
 The source node $s$ prepares an input quantum state $\ket{\psi}_A$ on system $A$ 
and makes $1\rightarrow 2$ asymmetric universal clones by applying an isometry $U_{1\rightarrow2}^{(a,b)}$ defined by Eq.~(\ref{1+1}) from the system $A$ to the system $ABM$.

\STATE \textbf{Step 2:} 
The source node $s$ measures  system $M$ in the computational basis, and sends the measurement outcome $r$ to the two target nodes $t_1$ and $t_2$.

\STATE \textbf{Step 3:} 
The source node $s$ applies the unitary operator $\Upsilon _{AB}$ defined by Eqs.(\ref{eq def upsilon 1}) and (\ref{1+1clone}) to the systems $AB$, then discards the system $B$. 

\STATE \textbf{Step 4:} 
The state on  system $A$ is multicast to the target nodes $t_1$ and $t_2$ over the quantum network $G$ using Kobayashi et al.'s protocol. 
The target nodes $t_1$ and $t_2$ put the output GHZ-type state of Kobayashi et al.'s protocol on system $CD$.

\STATE \textbf{Step 5:} 
The target nodes $t_1$ and $t_2$ apply $X_d^{r-1} \otimes X_d^{r-1}$ to system $EF$ using the measurement outcome $r$ sent from the source node $s$.

\STATE \textbf{Step 6:} 
The target node $t_1$ applies $V_{C,E}^{(r)}$ defined by Eq.~(\ref{eq def V r}) to system $CE$,
and the target node $t_2$ applies $V_{D,F}^{(r)}$ to system $DF$ . Then, The target node $t_1$ applies $\Delta_{C,E}^{(r)}$ defined by Eq.~(\ref{eq def Delta r}) to system $CE$,
and the target node $t_2$ applies  $\Delta_{D,F}^{(r)}$ to system $DF$.

\STATE \textbf{Step 7:} 
The target node $t_1$ initializes $G$ in $\ket{0}$, and applies $\Gamma_{C,E,G}^{(r)}$ defined by Eq.~(\ref{eq def Gamma r}) on system $CEG$. The target node $t_2$ initializes 
$H$ in $\ket{0}$ and applies $\Gamma_{C,E,G}^{(r)}$ to system  $DFH$.

\STATE \textbf{Step 8:} 
The target node $t_2$ sends the state on system $H$ to system $T_1$ at the target node $t_1$ using the Bell state on system $T_1T_2$ by the quantum teleportation.

\STATE \textbf{Step 9:} 
The target node $t_1$ applies $\Theta_{GT_1}$ defined by Eq.~(\ref{step9}) to systems $G$ and $T_1$, and discards $T_ 1$.

\STATE \textbf{Step 10:} 
The target node $t_1$ measures system $G$ in \[\left \{|\tilde{0} \rangle = \frac{|0\rangle + |1\rangle}{\sqrt{2}},|\tilde{1} \rangle = \frac{|0\rangle - |1\rangle}{\sqrt{2}} \right \}\] 
and derives the measurement outcome $k$. 
Then, $t_1$ performs $Z_{2}^{-k}$ on system $G$.

\STATE \textbf{Step 11:} 
The target node $t_1$ applies  $\Lambda^{(r)}_{CE} $ defined by Eq.~(\ref{eq def lambda}) on the system $CE$, 
and the target node $t_2$ applies $ \Lambda^{(r)}_{DF}$ the system $DF$.

\STATE \textbf{Step 12:} 
The target nodes $t_1$ and $t_2$ measure system $C$ and $D$ in the Fourier basis 
\[ \left \{\sum_{x=0}^{d-1}\frac{\omega^{px}}{\sqrt{d}}|x\rangle \right \}_{p \in Z_d}, \] respectively, and derive the measurement outcomes $p_1$ and $p_2$, respectively.
Then, $t_1$ applies $Z^{p_1}$ to system $E$, and $t_2$ applies $Z^{p_2}$ to system $F$ .

\end{algorithmic}
\end{Protocol}

The protocol for $1\rightarrow 2$ is shown as  protocol$2$.
Using the protocol, $1\rightarrow 2$ asymmetric UQCs given by Eq.~(\ref{eq asymmetric uqcm}) are created in systems $EF$, where $E$ and $F$ are on the target nodes $t_1$ and $t_2$, respectively.
Note that as we explained in the previous section, asymmetric UQCs depend on the parameters $a$ and $b$ in Eq.~(\ref{1+1}). 
We can set these parameters in step 2 of the protocol, when we apply $U_{1\rightarrow 2}^{(a,b)}$.

\subsection{Proof of $1\rightarrow 2$\ quantum multicast protocol}\label{subsec proof 1 2}
In this section, we present the proof that protocol 2 creates $1\rightarrow 2$ asymmetric UQCs given by Eq.~(\ref{eq asymmetric uqcm}) in system $EF$.

As we explained in the previous section, an input state at the source node $s$ can be written as 
\[ \ket{\psi} = \sum_{j=0}^{d-1} \alpha_j |j \rangle \in \mathcal{H}_s.\] 
Then, from Eq.~(\ref{1+1}), the state on system $ABM$ after step 1 can be written as:
\begin{eqnarray}
a|\psi \rangle_A|\Phi^+\rangle_{BM} + b|\psi \rangle_B|\Phi^+\rangle_{AM}
\end{eqnarray}

The unnormalized state $\ket{\Psi_2^{(r)}}_{AB}$ on system $AB$ after deriving measurement outcome $r$ in Step 2 can be written as:
\begin{eqnarray}
\ket{\Psi_2^{(r)}}_{AB} := \beta _r \ket{rr}_{AB}+ \sum_{j\neq r} \beta_j
\left( \cos \eta \ket{jr}_{AB} + \sin \eta \ket{rj}_{AB} \right) \label{step2},
\end{eqnarray}
where $\eta$ is defined by Eq.~(\ref{eq def eta}), and  $\left \{ \beta_j \right \}_{j=0}^{d-1}$ is defined by 
\begin{align}
\beta_r & =\frac{\alpha_r}{\sqrt{d}}(a+b) \nonumber \\
\beta_j & =\frac{\alpha_j}{\sqrt{d}}\sqrt{1-\frac{2ab}{d}} \quad (\forall j \neq r).
\end{align}
Here, $\left \| \ket{\Psi_2^{(r)}}_{AB} \right \| ^2= \sum _j  |\beta _j|^2$ is a probability in which outcome $r$ is derived in step 2. 
Since measuring  system $M$ without seeing the outcome is mathematically equivalent to tracing out system $M$,  $\{ \ket{\Psi_2^{(r)}}\}_{r=0}^{d-1}$ satisfies 
\begin{equation}\label{eq varepsilon psi Psi 2}
\varepsilon_{1\rightarrow 2}\left( \ket{\psi}\bra{\psi} \right) = \sum_{r=0}^{d-1} \ket{\Psi_2^{(r)}}\bra{\Psi_2^{(r)}},
\end{equation}
where $\varepsilon_{1\rightarrow 2}$ is a $1\rightarrow 2$ optimal asymmetric UQCM defined by Eq.~(\ref{eq asymmetric uqcm}).
Hence, the purpose of the remaining part of the protocol is to transfer $\ket{\Psi_2}$ to the target nodes.
However, in our problem settings, the throughput of the quantum network is too small to send $\ket{\Psi_2}$ directly to the target nodes.
Hence, first, we compress the state on the $d$-dimensional system in step 3. 
Then, the unnormalized state of system $AB$ after step 3 can be written as 
\begin{equation}
|\Psi_{3} \rangle_A = \sum_{j=0}^{d-1}\beta_j|j \rangle_A.
\end{equation}

In step 4, Kobayashi et al.'s protocol successfully works based on the assumption for the existence of a classical linear multicast network code. Since the (unnormalized) input state for Kobayashi et al.'s protocol is $\ket{\Psi_3}_A$, the unnormalized state on the system $C$ at the target node $t_1$ and on system $D$ at the target node $t_2$ can be written as  $\sum_{j=0}^{d-1}\beta_j|j \rangle_{C} |j \rangle_{D}$. The purpose of the remaining part of the protocol is to reconstruct $\ket{\Psi_2}$ from this state.

Since system $EF$ is initially on $\cos \eta |0\rangle_E |1 \rangle_E + \sin \eta|1 \rangle_E |0 \rangle_F$, the unnormalized  state on system $CDEF$ can be written as 
\begin{eqnarray}
\sum_{j=0}^{d-1} \beta_j |jj \rangle_{CD} \otimes (\cos\eta |0 \rangle_E |1 \rangle_F + \sin\eta |1 \rangle_E |0 \rangle_F) 
\end{eqnarray}
Then, the  unnormalized  state on $CDEF$ after step 5 can be written as 
\begin{eqnarray}
\sum_{j=0}^{d-1} \beta_j |jj \rangle_{CD} \otimes (\cos\eta |r-1 \rangle_E |r \rangle_F + \sin\eta |r \rangle_E |r-1 \rangle_F).
\end{eqnarray}
The  unnormalized  state on $CDEF$ after step6 is
\begin{eqnarray}
\sum_{j \neq r} \beta_j |jj \rangle_{CD} \otimes (\cos\eta |j \rangle_E |r \rangle_F + \sin\eta |r \rangle_E |j \rangle_F) \nonumber \\
+ \beta_r |rr \rangle_{CD}  \otimes (\cos\eta |0 \rangle_E |1 \rangle_F + \sin\eta |1 \rangle_E |0 \rangle_F). 
\end{eqnarray}
Then, the  unnormalized  state on $CDEFGH$ after step 7 can be written  as 
\begin{eqnarray}
\sum_{j \neq r} \beta_j |jj \rangle_{CD} \otimes (\cos\eta |j \rangle_E |r \rangle_F + \sin\eta |r \rangle_E |j \rangle_F) \otimes |00 \rangle_{GH} \nonumber \\
+ \beta_r |rr \rangle_{CD} \otimes |00 \rangle_{EF} \otimes (\cos\eta |0 \rangle_G |1 \rangle_H + \sin\eta |1 \rangle_G |0 \rangle_H). 
\end{eqnarray}
Next, in step 8, the state on the system $H$ is transferred to system $T_1$ by quantum teleportation. Thus, the unnormalized state on $CDEFG$ after step 9 can be written as
\begin{eqnarray} 
\sum_{j \neq r} \beta_j |jj \rangle_{CD} \otimes (\cos\eta |j \rangle_E |r \rangle_F + \sin\eta |r \rangle_E |j \rangle_F) \otimes |0 \rangle_G \nonumber \\
+  \beta_r |rr \rangle_{CD} \otimes |00 \rangle_{EF} \otimes |1 \rangle_G.
\end{eqnarray}
Since system $G$ is effectively removed in step10, the unnormalized state on  system $CDEF$ after step 10 can be written as
\begin{align}
\sum_{j \neq r} & \beta_j |jj \rangle_{CD} \otimes (\cos\eta |j \rangle_E |r \rangle_F 
+ \sin\eta |r \rangle_E |j \rangle_F)
\nonumber \\
&\qquad  + \beta_r |rr \rangle_{CD} \otimes |00 \rangle_{EF}.
\end{align}
Then, the unnormalized state on $CDEF$ after step 11 can be written as
\begin{align}
\sum_{j \neq r} & \beta_j |jj \rangle_{CD} \otimes (\cos\eta |j \rangle_E |r \rangle_F + \sin\eta |r \rangle_E |j \rangle_F) \nonumber \\
& \qquad + \beta_r |rr \rangle_{CD} \otimes |rr \rangle_{EF}.
\end{align}

In step 12, after system $CD$ is measured in the Fourier basis 
$\{d^{-1/2} \cdot \sum_{x=0}^{d-1} \omega^{px}  |x\rangle \}_{p \in Z_d} $ and is discarded, the unnormalized state on $EF$ for the measurement outcomes $p_1$ and $p_2$ can be written as 
\begin{align}
\sum_{j \neq r} & \beta_j \omega^{-j(p_1+p_2)} (\cos\eta |j \rangle_E |r \rangle_F + \sin\eta |r \rangle_E |j \rangle_F) \nonumber \\
& + \beta_r \omega^{-r(p_1+p_2)} |rr\rangle_{EF} 
\end{align}
Hence, after applying $Z^{p_1} \otimes Z^{p_2}$ on system $EF$, the unnormalized state on $EF$ becomes
\begin{eqnarray}
\sum_{j=0,j \neq r}^{d-1} \beta_j  (\cos\eta |j \rangle_E |r \rangle_F + \sin\eta |r \rangle_E |j \rangle_F) + \beta_r |rr\rangle_{EF}. \label{step12}
\end{eqnarray}
This state is the state $\ket{\Psi _2^{(r)}}$ defined by Eq.~(\ref{step12}).
Since Eq.~(\ref{step12}) is the unnormalized state corresponding to the outcome $r$ in step 2,  the final state of this protocol can be written as $\sum_r \ket{\Psi _2^{(r)}}\bra{\Psi _2^{(r)}}$. Hence, by Eq.~(\ref{eq varepsilon psi Psi 2}), the final states of protocol 2 on the target nodes $t_1$ and $t_2$ are $1\rightarrow 2$ optimal asymmetric UQCs of the input state $\ket{\psi}$.

\section{$1\rightarrow 3$ optimal asymmetric quantum universal clones multicast protocol}\label{sec:iv}
In this section, we present a protocol that multicasts optimal asymmetric UQCs from the source node $s$ to three target nodes $t_1$, $t_2$ and $t_3$ on a quantum network.  We present the protocol in Section~\ref{protocol 1 3} and in Section~\ref{subsec proof 13}, we prove that creates optimal asymmetric UQCs.

\subsection{$1\rightarrow 3$\ quantum multicast protocol}\label{protocol 1 3}
In this section, we present a protocol that multicasts $1\rightarrow 3$ optimal asymmetric UQCs of an input quantum state  from  the source node $s$ to two target nodes $t_1$, $t_2$ and $t_3$.  

The problem setting for the $1\rightarrow 3$ quantum multicast protocol is almost the same as that of the $1\rightarrow 2$ protocol given in the last section. 
Hence, we consider only  the difference between these two problem settings.
First, the number of target nodes is different. That is, 
in this section, a quantum network $G$ has three target nodes $t_1, t_2$, and $t_3$.
The purpose of the protocol is to construct $1\rightarrow 3$ optimal asymmetric universal clones given by Eq.~(\ref{eq def asymmetric uqcm 13}) among target nodes $t_1, t_2$ and $t_3$ for a given $d$-dimensional input state $\ket{\psi}=\sum _{j=0}^{d} \alpha_j \ket{j} \in \mathcal{H}_s$ on a source node, where $\mathcal{H}_s$ is a $d$-dimensional input space.
We again assume $d=q^r$. In other words, we consider a mulcast quantum network code with source rate $r$. The assumption for the existence of a classical linear multicast network code is also similar. That is, a classical linear multicast network code is 
a code on $\mathbb{F}_q$ used to multicast from the node $s$ to the nodes $t_1, t_2, t_3$ on $G'$ with source rate $r$. The amount of entanglement shared among the target nodes is also different. In $1 \rightarrow 3$ case, we assume that at most $2+4 \log _2 3$\,ebits are shared among the target nodes $t_1, t_2$ and $t_3$. Hence, the amount of this entanglement resource is constant with respect to the dimension $d$ of the input state.   

Before we present the protocol, we define the unitary operators used in the protocol. 
$U_2^{(r,s)}$ is a tripartite unitary operator satisfying the following conditions:
\begin{align}
&  U_2^{(r,s)} \cdot \frac{\alpha|jrs\rangle + \beta|rjs\rangle + \gamma|rsj\rangle + \alpha|jsr\rangle + \beta|sjr\rangle + \gamma|srj\rangle  }{\sqrt{2\alpha^{2}+2\beta^{2}+2\gamma^{2}}} & \nonumber \\
 & \qquad \qquad  \qquad \qquad \qquad \qquad \qquad  = |j00\rangle, \quad  (\forall j \neq r,s) \nonumber 
\\
& U_2^{(r,s)} \cdot  \frac{\left (\alpha+\beta \right )|rrs\rangle + \left (\beta+\gamma \right )|srr\rangle + \left (\gamma+\alpha \right )|rsr\rangle}{\sqrt{(\alpha+\beta)^{2}+(\beta+\gamma)^{2}+(\gamma+\alpha)^{2}}}  =|r00\rangle,\nonumber   
\\
& U_2^{(r,s)} \cdot \frac{(\alpha+\beta)|ssr\rangle + (\beta+\gamma)|rss\rangle + (\gamma+\alpha)|srs\rangle}{\sqrt{(\alpha+\beta)^{2}+(\beta+\gamma)^{2}+(\gamma+\alpha)^{2}}}  = |s00\rangle. \label{eq def U 2 rs}
\end{align}
$U_5^{(r,s)}$ is a bipartite unitary operator defined by 
\begin{equation}\label{eq def U 5 rs}
U_5^{(r,s)} := |r\rangle \langle r| \otimes I + |s\rangle \langle s| \otimes I + \sum_{j \neq r, s}^{d-1}|j \rangle \langle j| \otimes (\sum_{x=0}^{d-1}|\pi_{jrs}(x)\rangle \langle x|),
\end{equation} 
where $\pi_{jrs}$ is a permutation satisfying the following conditions:
\begin{equation}
\pi_{jrs}(0) = j, \quad \pi_{jrs}(1) = r, \quad \pi_{jrs}(2) = s 
\end{equation}
$U_6^{(r,s)}$ is a tripartite unitary operator defined by 
\begin{equation}\label{eq def U 6 rs}
U_6^{(r,s)}:=|r\rangle \langle r|\otimes swap + |s\rangle \langle s|\otimes swap + \sum_{j \neq r, s} |j \rangle \langle j| \otimes I \otimes I,
\end{equation} 
where $swap$ is a swap operator defined by $swap:=\sum_{i,j=0}^{d-1}|ij \rangle \langle ji|$.
$U_7^{(r,s)}$ is a bipartite unitary operator defined by 
\begin{equation}\label{eq def U 7 rs}
U_7^{(r,s)} := \sum_{i \neq r, s} |i\rangle \langle i| \otimes I + |r\rangle \langle r| \otimes \sum_{j=0}^{d-1}|\pi_{rs}^{\prime \prime}(j) \rangle \langle j| + |s \rangle \langle s|\otimes \sum_{k=0}^{d-1}|\pi_{sr}^{\prime \prime}(k) \rangle \langle k|,
\end{equation} 
where $\pi_{xy}^{\prime \prime}$ is a permutation satisfying $\pi_{xy}^{\prime \prime}(0)=x$ and $\pi_{xy}^{\prime \prime}(1)=y$. 
$U_8$ is a unitary operator on $\mathbb{C}^3 \otimes \mathbb{C}^3 \otimes \mathbb{C}^3$ satisfying
\begin{align}
& U_8(\alpha_1^{\prime \prime}|001\rangle + \beta_1^{\prime \prime}|100\rangle + \gamma_1^{\prime \prime}|010\rangle) =|000\rangle \nonumber \\
& U_8 (\alpha_1^{\prime}( |012\rangle+ |021\rangle) 
+ \beta_1^{\prime}(|102\rangle +|201\rangle) 
+ \gamma_1^{\prime}( |120\rangle  + |210\rangle))\nonumber \\
&\qquad \qquad \qquad \qquad \qquad \qquad = |100\rangle, \label{eq def U 8 rs}
\end{align}
where $\alpha_1^\prime$, $\beta_1^\prime$, $\gamma_1^\prime$, $\alpha_1^{\prime \prime}$, $\beta_1^{\prime \prime}$, and  $\gamma_1^{\prime \prime}$ are defined by 
\begin{align} \label{eq alpha beta gamma prime}
& \alpha_1^{\prime} = \frac{\alpha}{\sqrt{2\alpha^{2} + 2\beta^{2} + 2\gamma^{2}}}, \quad \beta_1^{\prime} = \frac{\beta}{\sqrt{2\alpha^{2} + 2\beta^{2} + 2\gamma^{2}}},\nonumber \\
& \gamma_1^{\prime} = \frac{\gamma}{\sqrt{2\alpha^{2} + 2\beta^{2} + 2\gamma^{2}}} \nonumber \\
& \alpha_1^{\prime \prime} = \frac{\alpha + \beta}{\sqrt{(\alpha + \beta)^{2} + (\beta + \gamma)^{2} + (\gamma  +\alpha)^{2}}}, \nonumber \\
& \beta_1^{\prime \prime} = \frac{\beta + \gamma}{\sqrt{(\alpha + \beta)^{2} + (\beta + \gamma)^{2} + (\gamma  +\alpha)^{2}}}, \nonumber \\
& \gamma_1^{\prime \prime} = \frac{\gamma + \alpha}{\sqrt{(\alpha + \beta)^{2} + (\beta + \gamma)^{2} + (\gamma  +\alpha)^{2}}}.
\end{align}
$U_{9}^{(r,s,k)}$ is a unitary operator on $\mathbb{C}^d$ defined by
\begin{equation}\label{eq def U 9 rs}
U_9^{(r,s,k)} := \sum_{j \neq r, s}|j\rangle \langle j| + (-1)^{k}|r\rangle \langle r| +  (-1)^{k}|s\rangle \langle s|.
\end{equation}

$U_2^{\prime(r)}$ is a tripartite unitary operator satisfying
\begin{align}
U_2^{\prime(r)}\cdot \frac{2\alpha|jrr\rangle + 2\beta|rjr\rangle + 2\gamma|rrj\rangle}{\sqrt{(2\alpha)^{2}+(2\beta)^{2}+(2\gamma)^{2}}} & = |j00\rangle, \quad  (\forall j \neq r ) \nonumber \\
U_2^{\prime(r)} |rrr\rangle & = |r00\rangle \label{eq def U 2 r}
\end{align}
$U_5^{\prime(r)}$ is a bipartite unitary operator defined by 
\begin{equation}\label{eq def U 5 r}
U_5^{\prime(r)} := |r\rangle \langle r| \otimes I + \sum_{j \neq r} |j \rangle \langle j| \otimes \left ( \sum_{x=0}^{d-1}|\pi_{jr}(x)\rangle \langle x| \right ),
\end{equation}
where $\pi_{jr}$ is a permutation satisfying 
\begin{equation}
\pi_{jr}(1) = r, \quad 
\pi_{jr}(0) = j 
\end{equation} 
$U_6^{\prime(r)}$ is a tripartite unitary operator defined by 
\begin{equation}\label{eq def U 6 r}
U_6^{\prime(r)}:=|r\rangle \langle r |\otimes \operatorname{swap}+ \sum_{ j \neq r}^{d-1}|j \rangle \langle j| \otimes I \otimes I,
\end{equation}
where $\operatorname{swap}$ is an operator defined by $\operatorname{swap}:=\sum_{i,j=0}^{d-1}|ij \rangle \langle ji|$.
$U_7^{\prime(r)}$ is a tripartite unitary operator defined by 
\begin{equation}\label{eq def U 7 r}
U_7^{\prime(r)} := \sum_{i \neq r} |i\rangle \langle i| \otimes I + |r\rangle \langle r| \otimes X^{r},
\end{equation} 
where $X$ is the Pauli $X$ operator defined by $X :=\sum_{x=0}^{d-1}|x\oplus 1\rangle \langle x|$.
$U'_8$ is a unitary operator on $\mathbb{C}^2 \otimes \mathbb{C}^2 \otimes \mathbb{C}^2$ defined by 
\begin{align}
U'_8 |000\rangle &= |000\rangle \nonumber \\
U'_8 (\alpha_2^{\prime}|011\rangle + \beta_2^{\prime}|101\rangle + \gamma_2^{\prime}|110\rangle) &= |100\rangle \label{eq def U 8 r}
\end{align}
Finally, $U_{9}^{\prime (r,k)}$ is a unitary operator on $\mathbb{C}^d$ defined by
\begin{equation}\label{eq def U 9 r}
U_{9}^{\prime (r,k)} := \sum_{ j \neq r}|j\rangle \langle j| + (-1)^{k}|r\rangle \langle r|.
\end{equation}
We will also use in the protocol the projective measurement $\{P_k \}_{k=0}^2 $ defined by the following equations:
\begin{equation}\label{eq def P k}
P_0:= \ket{\tilde{0}}\bra{\tilde{0}}, P_1:=\ket{\tilde{1}}\bra{ \tilde{1}}, P_2:= I-\ket{\tilde{0}}\bra{ \tilde{0}} -\ket{\tilde{1}}\bra{ \tilde{1}},
\end{equation}
where $\ket{\tilde{0}}:=\frac{\ket{0}+\ket{1}}{\sqrt{2}}, \ket{\tilde{1}}:=\frac{\ket{0}-\ket{1}}{\sqrt{2}}$.

At the beginning of the protocol, the source node $s$ has five $d$-dimensional systems $A$, $B$, $C$, $R$ and $S$.
The target node $t_1$ has three $d$-dimensional systems $D$, $M_1$, and $N_1$.
The target node $t_2$ has three $d$-dimensional systems $E$, $M_2$, and $N_2$.
The target node $t_3$ has three $d$-dimensional systems $F$, $M_3$, and $N_3$.
Further, the target nodes $t_1$ and $t_2$ share 
$1+2\log_2 3$\,ebits of entanglement, 
and the target nodes $t_1$ and $t_2$ share 
$1+2\log_2 3$\,ebits of entanglement. 
Hence, the amount of entanglement resources are $2+4\log _2 3$\,ebits in total.

\begin{Protocol}
\caption{$1\rightarrow 3$\ quantum multicast network coding protocol (beginning)}         
\begin{algorithmic}
\STATE \textbf{Step 1:} 
 The source node $s$ prepares an input quantum state $\ket{\psi}_A$ on system $A$, 
and makes $1\rightarrow 3$ asymmetric universal clones by applying an isometry $U_{1\rightarrow3}^{(\alpha,\beta,\gamma)}$ defined by Eq.~(\ref{1+1+1}) from system $A$ to system $ABCRS$.

\STATE \textbf{Step 2:} 
The source node $s$ measures the systems $R$ and $S$ in the computational basis, where the measurement outcomes of $R$ and $S$ are called $r$ and $s$, respectively. The source node $s$ sends the measurement outcomes $r$ and $s$ to the target nodes $t_1$ $t_2$, and $t_3$.
The following steps of the protocol depend on whether $r \neq  s$ or $r=s$.
\end{algorithmic}
\end{Protocol}

\begin{Protocol}
\caption{Continuation of protocol 3 for $1\rightarrow 3$\ quantum multicast network coding (for $r\neq s$)}         
\begin{algorithmic}
\STATE \textbf{[$r\neq s$]}

\STATE \textbf{Step 3:} 
The source node $s$ applies unitary operator $U_2^{(r,s)}$ defined by Eq.~(\ref{eq def U 2 rs}) to  system $ABC$, and then, discards systems $B$ and $C$.

\STATE \textbf{Step 4:} 
The state on system $A$ is multicast to the target nodes $t_1$, $t_2$ and $t_3$ over the quantum network $G$ using Kobayashi et al.'s protocol. 
The target nodes $t_1$, $t_2$ and $t_3$ put the output of Kobayashi et al.'s protocol on system $DEF$.
Then, using $2 \log_2 3$\,ebits of entanglement, the targets nodes share 
the following state on  system $M_1M_2M_3$:
\begin{eqnarray*}
\left ( \alpha_1^{\prime}( |012\rangle+ |021\rangle)  
+ \beta_1^{\prime}(|102\rangle + |201\rangle)
+ \gamma_1^{\prime}(|120\rangle  +|210\rangle )\right )_{M_1,M_2,M_3},
\end{eqnarray*}
where $\alpha_1^{\prime}$, $\beta_1^{\prime}$, and $\gamma_1^{\prime}$ are defined by Eq.~(\ref{eq alpha beta gamma prime}).
Further, by using $2$\,ebits of entanglement, the target nodes share the following state on system $N_1N_2N_3$:
\begin{eqnarray}
(\alpha_1^{\prime \prime}|001\rangle + \beta_1^{\prime \prime}|100\rangle + \gamma_1^{\prime \prime}|010\rangle)_{N_1,N_2,N_3},
\end{eqnarray}
where $\alpha_1^{\prime \prime}$, $\beta_1^{\prime \prime}$ and $\gamma_1^{\prime \prime}$ are defined by Eq.~(\ref{eq alpha beta gamma prime}).

\STATE \textbf{Step 5:} 
The target nodes apply $U_{5,DM_1}^{(r,s)} \otimes U_{5, EM_2}^{(r,s)} \otimes U_{5,FM_3}^{(r,s)}$  to system $DM_1EM_2FM_3$, where $ U_5^{(r,s)}$ is defined by Eq.~(\ref{eq def U 5 rs}).
\STATE \textbf{Step 6:} 
The target nodes apply  $U_{6,DM_1N_1}^{(r,s)} \otimes U_{6,EM_2N_2}^{(r,s)} \otimes U_{6,FM_3N_3}^{(r,s)}$ to  system $DM_1N_1EM_2N_2FM_3N_3$, where $ U_6^{(r,s)}$ is defined by Eq.~(\ref{eq def U 6 rs}).

\STATE \textbf{Step 7:} 
The target nodes apply $U_{7,DM_1}^{(r,s)} \otimes U_{7,EM_2}^{(r,s)} \otimes U_{7,FM_3}^{(r,s)}$ to system $DM_1EM_2FM_3$, where $ U_7^{(r,s)}$ is defined by Eq.~(\ref{eq def U 7 rs}).

\STATE \textbf{Step 8:} 
Using $2\log_2 3$\,ebits of entanglement resource, 
subspaces spanned by $\{|0\rangle,|1\rangle,|2\rangle\}$ of the systems $N_2$ and $N_3$ are sent from the target nodes $t_2$ and $t_3$ to the target node $t_1$, respectively. 
The target node $t_1$ applies $U_{8, N_1N_2N_3}$ to system $N_1N_2N_3$  and discards systems $N_2$ and $N_3$.

\STATE \textbf{Step 9:} 
The target node $t_1$ applies the projective measurement  $\left \{ P_k \right \}_{k=0}^2$ defined by Eq.~(\ref{eq def P k}) on system $N_1$ in the basis  and discards the quantum system $N_1$. 
Then, depending on the measurement outcome $k$, the target node $t_1$ applies $U_9^{(r,s,k)}$ defined by Eq.~(\ref{eq def U 9 rs}) on system $D$.

\STATE \textbf{Step 10:} 
The target nodes $t_1$, $t_2$ and $t_3$ measure system $D$, $E$ and $F$ in the Fourier basis $\{d^{-1/2} \cdot \sum_{x=0}^{d-1}\omega^{px}|x\rangle \}_{p \in \mathbb{Z}_d}$, respectively.  Then,
they apply $Z^{(p_1+p_2+p_3)} \otimes Z^{(p_1+p_2+p_3)} \otimes Z^{(p_1+p_2+p_3)}$ to system $M_1M_2M_3$, where $p_1$, $p_2$ and $p_3$ are the measurement outcomes on the target nodes $t_1$, $t_2$ and $t_3$, respectively.
\end{algorithmic}
\end{Protocol}

\begin{Protocol}
\caption{Continuation of protocol 3 for $1\rightarrow 3$\ quantum multicast network coding (for $r = s$)}         
\begin{algorithmic}
\STATE \textbf{[$r = s$]}
\STATE \textbf{Step 3:} 
The source node $s$ applies unitary operator $U_2^{\prime(r)}$ defined by Eq.~(\ref{eq def U 2 r}) to system $ABC$, and then, discards the systems $B$ and $C$.

\STATE \textbf{Step 4:} 
The state on system $A$ is multicast to the target nodes $t_1$, $t_2$ and $t_3$ over the quantum network $G$ using Kobayashi et al.'s protocol. 
The target nodes $t_1$, $t_2$ and $t_3$ put the output of Kobayashi et al.'s protocol on system $DEF$.
Then, using $2$\,ebits of entanglement, the target nodes share the following state on system $M_1M_2M_3$:
\begin{eqnarray}
\left (\alpha_2^{\prime}|011\rangle + \beta_2^{\prime}|101\rangle + \gamma_2^{\prime}|110\rangle \right )_{M_1M_2M_3},
\end{eqnarray}
where $\alpha_2^{\prime} = \frac{2\alpha}{\sqrt{(2\alpha)^{2} + (2\beta)^{2} + (2\gamma)^{2}}}, \beta_2^{\prime} = \frac{2\beta}{\sqrt{(2\alpha)^{2} + (2\beta)^{2} + (2\gamma)^{2}}}$ and $\gamma_2^{\prime} = \frac{2\gamma}{\sqrt{(2\alpha)^{2} + (2\beta)^{2} + (2\gamma)^{2}}}$.\\
Further, they initialize all the systems $N_1$, $N_2$, and $N_3$ in $\ket{0}$.

\STATE \textbf{Step 5:} 
The target nodes apply $U_{5,DM_1}^{\prime(r)} \otimes U_{5,EM_2}^{\prime(r)} \otimes U_{5,FM_3}^{\prime(r)}$ to system $DM_1EM_2FM_3$, where $U_5^{\prime(r)}$ is defined by Eq.~(\ref{eq def U 5 r}).
\STATE \textbf{Step 6:} 
The target nodes apply $U_{6,DM_1N_1}^{\prime(r)} \otimes U_{6,EM_2N_2}^{\prime(r)} \otimes U_{6,FM_3N_3}^{\prime(r)}$ to system $DM_1N_1EM_2N_2FM_3N_3$, where $U_6^{\prime(r)}$ is defined by Eq.~(\ref{eq def U 6 r}).

\STATE \textbf{Step 7:} 
The target nodes apply  $U_{7,DM_1}^{\prime(r)} \otimes U_{7,EM_2}^{\prime(r)} \otimes U_{7,FM_3}^{\prime (r)}$ to system $DM_1EM_2FM_3$, where $ U_7^{\prime(r)}$ is defined by Eq.~(\ref{eq def U 7 r}).
\STATE \textbf{Step 8:} 
By using $2$\,ebits of entanglement resource, 
subspaces spanned by $\{|0\rangle, |1\rangle \}$ of the systems $N_2$ and $N_3$ are sent from the target nodes $t_2$ and $t_3$ to the target node $t_1$, respectively. 
The target node $t_1$ applies $U'_{8, N_1N_2N_3}$ as defined by Eq.~(\ref{eq def U 8 r}) to system $N_1N_2N_3$  and discards system $N_2$ and $N_3$.

\STATE \textbf{Step 9:} 
The target node $t_1$ applies the projective measurement  $\left \{ P_k \right \}_{k=0}^2$ defined by Eq.~(\ref{eq def P k}) on system $N_1$ in the basis  and discards the quantum system $N_1$. 
Then, depending on the measurement outcome $k$, the target node $t_1$ applies $U_9^{(r,k)}$ defined by Eq.~(\ref{eq def U 9 r}) on the system $D$.

\STATE \textbf{Step 10:} 
The target nodes $t_1, t_2$ and $t_3$ measure system $D$, $E$ and $F$ in the Fourier basis $\{d^{-1/2}\cdot \sum_{x=0}^{d-1}\omega^{px}|x\rangle \}_{p \in \mathbb{Z}_d}$, respectively.  Then,
they apply $Z^{(p_1+p_2+p_3)} \otimes Z^{(p_1+p_2+p_3)} \otimes Z^{(p_1+p_2+p_3)}$ to system $M_1M_2M_3$, where $p_1$, $p_2$, and $p_3$ are the measurement outcomes on the target nodes $t_1, t_2$ and $t_3$, respectively.
\end{algorithmic}
\end{Protocol}

The beginning of the protocol for $1\rightarrow 3$ is given as protocol $3$.
In step 2 of protocol $3$, the systems $R$ and $S$ are measured and the measurement outcomes $r$ and $s$ are derived. 
The continuation of the protocol branches depending on whether $r \neq s$ or $r=s$. The continuation for  $r \neq s$ is given as protocol 4, and 
for $r=s$ is given as protocol 5. 
Using the protocol, $1\rightarrow 3$ asymmetric UQCs given by Eq. (\ref{eq def asymmetric uqcm 13}) are created system $M_1M_2M_3$, where $M_1$, $M_2$, and $M_3$ are on the target nodes $t_1$, $t_2$ and $t_3$, respectively.
Note that as we explained in the previous section, asymmetric UQCs depends on the parameters $\alpha$, $\beta$, and $\gamma$ in Eq.~(\ref{1+1+1}). 
We can set these parameters in step 1 of the protocol, when we apply $U_{1\rightarrow 3}^{(\alpha,\beta,\gamma)}$.

\subsection{Proof of $1\rightarrow 3$\ quantum multicast protocol}\label{subsec proof 13}
In this section, we prove that protocols 3, 4, and 5 create $1\rightarrow 3$ asymmetric UQCs given by Eq.~(\ref{eq def asymmetric uqcm 13})  in system $M_1M_2M_3$.

Let the input state at the source node be $|\psi \rangle = \sum_{j=0}^{d-1}\delta_j|j\rangle$. Then, from Eq.~(\ref{1+1+1}), the state on system $ABCRA$ can be written as  
\begin{align}
\sqrt{\frac{d}{2d+2}}[ & \alpha | \psi \rangle_A(|\Phi^{+} \rangle_{BR}|\Phi^{+} \rangle_{CS}+ |\Phi^{+} \rangle_{BS}|\Phi^{+} \rangle_{CR}) \nonumber \\
+ & \beta|\psi \rangle_B(|\Phi^{+}\rangle_{AR}|\Phi^{+}\rangle_{CS} + |\Phi^{+}\rangle_{AS}|\Phi^{+}\rangle_{CR})   \nonumber \\
+ & \gamma|\psi \rangle_C(|\Phi^{+} \rangle_{AR}|\Phi^{+} \rangle_{BS} + |\Phi^{+} \rangle_{AS}|\Phi^{+} \rangle_{BR})]
\end{align}
After step 2, the protocol branches depending on whether $r\neq s$ or $r=s$, where $r$ and $s$ are the measurement outcomes of system $R$ and $S$, respectively.

The unnormalized state $\ket{\Psi _2^{(r,s)}}$ after step2 for $r\neq s$ can be written as 
\begin{align}
&\ket{\Psi _2^{(r,s)}} \nonumber \\
=&\frac{1}{\sqrt{2d(d+1)}}\Big [\alpha (|\psi\rangle_A |r\rangle_B |s\rangle_C + |\psi \rangle_A |s\rangle_B |r\rangle_C) \nonumber \\
& + \beta (|r\rangle_A |\psi \rangle_B |s\rangle_C + |s \rangle_A |\psi\rangle_B |r\rangle_C) \nonumber \\
& \qquad + \gamma (|r \rangle_A |s \rangle_B |\psi \rangle_C + |s \rangle_A |r \rangle_B |\psi \rangle_C) \Big ] \nonumber \\
=& \frac{1}{\sqrt{2d(d+1)}}\Big[
\delta_r \big ( \left (\alpha + \beta  \right )|rrs\rangle \nonumber \\
& \quad \quad + \left (\beta + \gamma  \right )|srr\rangle + \left (\gamma + \alpha  \right )|rsr\rangle \big )_{ABC} \nonumber \\
&+ \delta_s \big ((\alpha + \beta)|ssr\rangle + (\beta + \gamma)|rss\rangle + (\gamma + \alpha)|srs\rangle \big )_{ABC}  \nonumber \\
&+ \sum _{j\neq r, s}
\delta_j \big (\alpha|jrs\rangle + \beta|rjs\rangle + \gamma|rsj\rangle \nonumber \\
& \qquad + \alpha|jsr\rangle + \beta|sjr\rangle + \gamma|srj\rangle \big)_{ABC}
 \Big ]. \label{rs}
\end{align} 
The unnormalized state $\ket{\Psi _2^{(r,r )}}$ after step 2 for $r= s$ can be written 
\begin{align}
 &\ket{\Psi _2^{ (r,r)}} \nonumber \\
=& \sqrt{\frac{2}{d(d+1)}}\left [\alpha|\psi\rangle_A |r \rangle_B |r \rangle_C + \beta |r \rangle_A |\psi \rangle_B |r \rangle_C + \gamma |r \rangle_A |r \rangle_B |\psi \rangle_C  \right ] 
\nonumber \\
=& \sqrt{\frac{2}{d(d+1)}} \Big [
 \delta_r(\alpha + \beta + \gamma)|rrr\rangle  + \nonumber \\
&\qquad \sum _{j \neq r} \delta_j(\alpha|jrr\rangle + \beta|rjr\rangle + \gamma|rrj\rangle) \Big ]. \label{r'}
\end{align}

As for the $1\rightarrow 2$ quantum multicast network coding protocol,  $\left \{ \ket{\Psi _2^{(r,s)}} \right \} _{r, s=0}^{d-1}$ satisfies 
\begin{equation}
\epsilon _{1\rightarrow 3}^{\alpha, \beta , \gamma} \left( \ket{\psi}\bra{\psi} \right) = \sum _{r, s=0}^{d-1}
 \ket{\Psi _2^{(r,s)}}  \bra{\Psi _2^{(r,s)}}, 
\end{equation}
where $\epsilon_{1 \rightarrow 3}^{\alpha, \beta , \gamma} $ is a $1 \rightarrow 3$ optimal asymmetric UQCM defined by Eq.~(\ref{eq def asymmetric uqcm 13}).
Hence, the purpose of the remaining part of the protocol is to transfer $\ket{\Psi _2^{(r,s)}}$ to the target nodes.

First we give the continuation  of the proof for $r\neq s$ (protocol 4).
We compress the state on a $d$-dimensional system on step 3.
The unnormalized state on system $A$ after step 3 can be written as
\begin{eqnarray}
\sum_{j=0}^{d-1}\kappa_j |j\rangle,
\end{eqnarray}
where $\{ \kappa_j \}_{j=0}^{d-1}$ is defined as 
\begin{align}
\kappa_j & = \sqrt{\frac{d}{2d+2}}\frac{\delta_j}{d}\sqrt{2\alpha^2 + 2\beta^2 + 2\gamma^2}\ \quad \quad (j \neq r, s), \nonumber \\
\kappa_{r} & = \sqrt{\frac{d}{2d+2}}\frac{\delta_{r}}{d}\sqrt{(\alpha+\beta)^2 + (\beta+\gamma)^2 + (\gamma+\alpha)^2}, \nonumber \\
\kappa_{s} & = \sqrt{\frac{d}{2d+2}}\frac{\delta_{s}}{d}\sqrt{(\alpha+\beta)^2 + (\beta+\gamma)^2 + (\gamma+\alpha)^2}.
\end{align}

In step 4, Kobayashi et al.'s protocol successfully works  based on the assumption for the existence of a classical linear multicast network code. The unnormalized state on system $D$ at the target node $t_1$, the system $E$ at the target node $t_2$, and on system $F$ at the target node $t_3$  can be written as  
\[\sum_{j=0}^{d-1}\kappa_j|j \rangle_{D} |j \rangle_{E}|j \rangle_{F}.\] 
Hence, the unnormalized state after step 4 can be written as 
\begin{align}
&\sum_{j=0}^{d-1}\kappa_j |jjj\rangle_{DEF}\otimes \Big (\alpha_1^{\prime}|012\rangle + \beta_1^{\prime}|102\rangle + \gamma_1^{\prime}|120\rangle \nonumber \\
&\qquad + \alpha_1^{\prime}|021\rangle + \beta_1^{\prime}|201\rangle + \gamma_1^{\prime}|210\rangle \Big )_{M_1M_2M_3} \nonumber \\
&\qquad \otimes (\alpha_1^{\prime \prime}|001\rangle + \beta_1^{\prime \prime}|100\rangle + \gamma_1^{\prime \prime}|010\rangle)_{N_1N_2N_3}
\end{align}
The purpose of the remaining part of the protocol is to reconstruct $\ket{\Psi _2^{(r,s)}} $ from the above  state. The unnormalized state after step 5 can be written as
\begin{widetext}
\begin{align}
& \Big ( \sum_{j \neq r,s}^{d-1} \kappa_j |jjj\rangle_{DEF}\otimes (\alpha_1^{\prime}|jrs\rangle + \beta_1^{\prime}|rjs\rangle + \gamma_1^{\prime}|rsj\rangle + \alpha_1^{\prime}|jsr\rangle + \beta_1^{\prime}|sjr\rangle + \gamma_1^{\prime}|srj\rangle)_{M_1M_2M_3} \nonumber \\
& + \left ( \kappa_r |rrr\rangle_{DEF} + \kappa_s |sss \rangle_{DEF} \right ) \otimes (\alpha_1^{\prime}|012\rangle + \beta_1^{\prime}|102\rangle + \gamma_1^{\prime}|120\rangle + \alpha_1^{\prime}|021\rangle + \beta_1^{\prime}|201\rangle + \gamma_1^{\prime}|210\rangle)_{M_1M_2M_3} \Big )  \nonumber \\
&\otimes (\alpha_1^{\prime \prime}|001\rangle + \beta_1^{\prime \prime}|100\rangle + \gamma_1^{\prime \prime}|010\rangle)_{N_1N_2N_3}
\end{align}
The unnormalized state after step 6 can be written as 
\begin{align}
& \sum_{j \neq r,s}^{d-1} \kappa_j |jjj\rangle_{DEF}\otimes (\alpha_1^{\prime}|jrs\rangle + \beta_1^{\prime}|rjs\rangle + \gamma_1^{\prime}|rsj\rangle+ \alpha_1^{\prime}|jsr\rangle + \beta_1^{\prime}|sjr\rangle + \gamma_1^{\prime}|srj\rangle)_{M_1M_2M_3} \otimes (\alpha_1^{\prime \prime}|001\rangle + \beta_1^{\prime \prime}|100\rangle + \gamma_1^{\prime \prime}|010\rangle)_{N_1N_2N_3}  \nonumber \\
& + \left(  \kappa_r |rrr\rangle  +\kappa_s |sss\rangle\right )_{DEF} \otimes (\alpha_1^{\prime \prime}|001\rangle + \beta_1^{\prime \prime}|100\rangle + \gamma_1^{\prime \prime}|010\rangle)_{M_1M_2M_3}  \nonumber \\
&\otimes (\alpha_1^{\prime}|012\rangle + \beta_1^{\prime}|102\rangle + \gamma_1^{\prime}|120\rangle + \alpha_1^{\prime}|021\rangle
+ \beta_1^{\prime}|201\rangle + \gamma_1^{\prime}|210\rangle)_{N_1N_2N_3}
\end{align}
Then, the unnormalized state after step 7 can be written as
\begin{align}
\sum_{j \neq r,s}^{d-1}& \kappa_j |jjj\rangle_{DEF}\otimes (\alpha_1^{\prime}|jrs\rangle + \beta_1^{\prime}|rjs\rangle + \gamma_1^{\prime}|rsj\rangle + \alpha_1^{\prime}|jsr\rangle + \beta_1^{\prime}|sjr\rangle + \gamma_1^{\prime}|srj\rangle)_{M_1M_2M_3} \otimes (\alpha_1^{\prime \prime}|001\rangle + \beta_1^{\prime \prime}|100\rangle + \gamma_1^{\prime \prime}|010\rangle)_{N_1N_2N_3}  \nonumber \\
+& \kappa_r |rrr\rangle_{DEF}\otimes (\alpha_1^{\prime \prime}|rrs\rangle + \beta_1^{\prime \prime}|srr\rangle + \gamma_1^{\prime \prime}|rsr\rangle)_{M_1M_2M_3} \otimes (\alpha_1^{\prime}|012\rangle + \beta_1^{\prime}|102\rangle + \gamma_1^{\prime}|120\rangle + \alpha_1^{\prime}|021\rangle 
+ \beta_1^{\prime}|201\rangle + \gamma_1^{\prime}|210\rangle)_{N_1N_2N_3}  \nonumber \\
+& \kappa_s |sss\rangle_{DEF} \otimes (\alpha_1^{\prime \prime}|ssr\rangle + \beta_1^{\prime \prime}|rss\rangle + \gamma_1^{\prime \prime}|srs\rangle)_{M_1M_2M_3}\otimes (\alpha_1^{\prime}|012\rangle + \beta_1^{\prime}|102\rangle + \gamma_1^{\prime}|120\rangle + \alpha_1^{\prime}|021\rangle
+ \beta_1^{\prime}|201\rangle + \gamma_1^{\prime}|210\rangle)_{N_1N_2N_3} 
\end{align}
The unnormalized state after step 8 can be written as 
\begin{align}
&\sum_{ j \neq r,s}^{d-1}\kappa_j |jjj\rangle_{DEF}\otimes (\alpha_1^{\prime}|jrs\rangle + \beta_1^{\prime}|rjs\rangle + \gamma_1^{\prime}|rsj\rangle + \alpha_1^{\prime}|jsr\rangle + \beta_1^{\prime}|sjr\rangle + \gamma_1^{\prime}|srj\rangle)_{M_1M_2M_3} \otimes |0\rangle_{N_1} \nonumber \\
&+\kappa_r |rrr\rangle_{DEF}\otimes (\alpha_1^{\prime \prime}|rrs\rangle + \beta_1^{\prime \prime}|srr\rangle + \gamma_1^{\prime \prime}|rsr\rangle)_{M_1M_2M_3} \otimes |1\rangle_{N_1} +\kappa_s |sss\rangle_{DEF} \otimes (\alpha_1^{\prime \prime}|ssr\rangle + \beta_1^{\prime \prime}|rss\rangle + \gamma_1^{\prime \prime}|srs\rangle)_{M_1M_2M_3}\otimes |1\rangle_{N_1} 
\end{align}
The unnormalized state after step 9 can be written as
\begin{align}
&\sum_{j \neq r,s}^{d-1}\kappa_j |jjj\rangle_{DEF}\otimes (\alpha_1^{\prime}|jrs\rangle + \beta_1^{\prime}|rjs\rangle + \gamma_1^{\prime}|rsj\rangle + \alpha_1^{\prime}|jsr\rangle + \beta_1^{\prime}|sjr\rangle + \gamma_1^{\prime}|srj\rangle)_{M_1M_2M_3} \nonumber \\
&+ \kappa_r |rrr\rangle_{DEF}\otimes (\alpha_1^{\prime \prime}|rrs\rangle + \beta_1^{\prime \prime}|srr\rangle + \gamma_1^{\prime \prime}|rsr\rangle)_{M_1M_2M_3} + \kappa_s |sss\rangle_{DEF} \otimes (\alpha_1^{\prime \prime}|ssr\rangle + \beta_1^{\prime \prime}|rss\rangle + \gamma_1^{\prime \prime}|srs\rangle)_{M_1M_2M_3}
\end{align}
The unnormalized state after step 10 can be written as 
\begin{align}
\omega^{(p_1+p_2+p_3)(r+s)}\Bigl\{&  \sum_{j \neq r,s}^{d-1}\kappa_j (\alpha_1^{\prime}|jrs\rangle + \beta_1^{\prime}|rjs\rangle + \gamma_1^{\prime}|rsj\rangle + \alpha_1^{\prime}|jsr\rangle + \beta_1^{\prime}|sjr\rangle + \gamma_1^{\prime}|srj\rangle)_{M_1M_2M_3} \nonumber \\
&+ \kappa_r (\alpha_1^{\prime \prime}|rrs\rangle + \beta_1^{\prime \prime}|srr\rangle + \gamma_1^{\prime \prime}|rsr\rangle)_{M_1M_2M_3} + \kappa_s (\alpha_1^{\prime \prime}|ssr\rangle + \beta_1^{\prime \prime}|rss\rangle + \gamma_1^{\prime \prime}|srs\rangle)_{M_1M_2M_3} \Bigr\} \label{proof_rs}
\end{align}
\end{widetext}

We can easily see that the above state is equivalent to $\ket{\Psi _2^{(r,s)}}$ as defined by Eq.~(\ref{rs})
except for a global phase. Hence, the proof is complete for $r \neq s$. 

Next, we give the continuation of the proof  for $r= s$ (protocol 5).
The unnormalized state on system $A$ after step 3 can be written as
\begin{eqnarray}
\sum_{j=0}^{d-1}\kappa_j^{\prime} |j\rangle,
\end{eqnarray}
where $\{ \kappa_j \}_{j=0}^{d-1}$ is defined as 
\begin{align}
\kappa_j^{\prime} & = \sqrt{\frac{2}{d(d+1)}}
\delta_j \sqrt{\alpha^2 + \beta^2 + \gamma^2}  \quad (j \neq r), \nonumber \\
 \kappa_{r}^{\prime} & = \sqrt{\frac{2}{d(d+1)}} \delta_{r}(\alpha+\beta + \gamma).
\end{align}
In step 4, Kobayashi et al.'s protocol successfully works, and the unnormalized state at the target nodes can be written as  $\sum_{j=0}^{d-1}\kappa_j|j \rangle_{D} |j \rangle_{E}|j \rangle_{F}$. Hence, the unnormalized state after step 4 can be written as 
\begin{align}
\sum_{j=0}^{d-1}\kappa_j^{\prime} |jjj\rangle_{DEF} & \otimes (\alpha_2^{\prime}|011\rangle + \beta_2^{\prime}|101\rangle + \gamma_2^{\prime}|110\rangle)_{M_1M_2M_3}\nonumber \\
& \otimes |000\rangle_{N_1N_2N_3}
\end{align}
Then, the unnormalized state after step 5 can be written as
\begin{align}
\sum_{j \neq r}^{d-1}& \kappa_j^{\prime} |jjj\rangle_{DEF}\nonumber \\
& \otimes (\alpha_2^{\prime}|jrr\rangle + \beta_2^{\prime}|r jr \rangle + \gamma_2^{\prime}|r r j\rangle)_{M_1M_2M_3}\otimes |000\rangle_{N_1N_2N_3} \nonumber \\
+ & \kappa_{r }^{\prime} |rrr\rangle_{DEF}\nonumber \\
& \otimes (\alpha_2^{\prime}|011\rangle + \beta_2^{\prime}|101\rangle + \gamma_2^{\prime}|110\rangle)_{M_1^{\prime}M_2^{\prime}M_3^{\prime}}\otimes |000\rangle_{N_1N_2N_3}
\end{align}
The unnormalized state after step 6 can be written as \begin{align}
\sum_{j \neq r}^{d-1}& \kappa_j^{\prime} |jjj\rangle_{DEF} \nonumber \\
& \otimes (\alpha_2^{\prime}|jrr\rangle + \beta_2^{\prime}|rjr\rangle + \gamma_2^{\prime}|rrj\rangle)_{M_1M_2M_3}\otimes |000\rangle_{N_1N_2N_3} \nonumber \\
+& \kappa_{r}^{\prime} |rrr\rangle_{DEF} \nonumber \\
&\otimes |000\rangle_{M_1M_2M_3}(\alpha_2^{\prime}|011\rangle + \beta_2^{\prime}|101\rangle + \gamma_2^{\prime}|110\rangle)_{N_1N_2N_3}
\end{align}
The unnormalized state after step 7 can be written as
\begin{align}
\sum_{j \neq r}^{d-1}& \kappa_j^{\prime} |jjj\rangle_{DEF}  \nonumber \\
& \otimes (\alpha_2^{\prime}|jrr\rangle + \beta_2^{\prime}|rjr\rangle + \gamma_2^{\prime}|rrj\rangle)_{M_1M_2M_3}\otimes |000\rangle_{N_1N_2N_3} \nonumber \\
+ & \kappa_j^{\prime} |rrr\rangle_{DEF} \nonumber \\
&\otimes |rrr\rangle_{M_1^{\prime}M_2^{\prime}M_3^{\prime}}(\alpha_2^{\prime}|011\rangle + \beta_2^{\prime}|101\rangle + \gamma_2^{\prime}|110\rangle)_{N_1N_2N_3}
\end{align}
Then, the unnormalized state after step 8 can be written as 
\begin{eqnarray}
\sum_{j \neq r}^{d-1}\kappa_j^{\prime} |jjj\rangle_{DEF}\otimes \Big (\alpha_2^{\prime}|jrr\rangle + \beta_2^{\prime}|rjr\rangle + \gamma_2^{\prime}|rrj\rangle)_{M_1M_2M_3}\otimes |0\rangle_{N_1} \nonumber \\
+ \kappa_{r}^{\prime} |rrr\rangle_{DEF}\otimes |rrr\rangle_{M_1^{\prime}M_2^{\prime}M_3^{\prime}}|1\rangle \Big )_{N_1}
\end{eqnarray}
The unnormalized state after step 9 can be written as 
\begin{eqnarray}
\sum_{j \neq r}^{d-1}\kappa_j^{\prime} |jjj\rangle_{DEF}\otimes (\alpha_2^{\prime}|jrr\rangle + \beta_2^{\prime}|rjr\rangle + \gamma_2^{\prime}|rrj\rangle)_{M_1M_2M_3} \nonumber \nonumber \\
+ \kappa_{r}^{\prime} |rrr\rangle_{DEF}\otimes |rrr\rangle_{M_1M_2M_3}
\end{eqnarray}
Finally, the unnormalized state after step 10 can be written as 
\begin{eqnarray}
\omega^{(p_1^{\prime}+p_2^{\prime}+p_3^{\prime})2r}\Bigl\{\sum_{j=0, j \neq r}^{d-1}\kappa_j^{\prime} (\alpha_2^{\prime}|jrr\rangle + \beta_2^{\prime}|rjr\rangle + \gamma_2^{\prime}|rrj\rangle)_{M_1M_2M_3} \nonumber \\
+ \kappa_{r}^{\prime} |rrr\rangle_{M_1M_2M_3} \Bigr\} \label{proof_r'}
\end{eqnarray}
We can easily see that the above state is equivalent to $\ket{\Psi _2^{(r,r)}}$ as defined by Eq.~(\ref{r'})
except faor  global phase. Hence, the proof is complete for $r = s$. 
Thus, we have achieved multicasting of asymmetric optimal clones for the systems $M_1$, $M_2$ and $M_3$ to the three target nodes.

\section{Conclusion}\label{sec:v}
In this paper, we considered quantum multicast network coding as the multicasting of optimal UQCs over a quantum network.
By extending Owari et al.'s results \cite{OKM13,KOM14,KOM15} for multicast of symmetric optimal UQCs,
we developed a protocol to multicast {\it asymmetric} optimal UQCs over a quantum network.  
Our results can be summarized as follows.
Suppose a quantum network is described by an undirected graph $G$ with one source node and two (three) target nodes, and each quantum channel on the quantum network $G$ can transmit one $q$-dimensional quantum system in a single session.
Further, suppose there exists a classical solvable multicast network code with source rate $r$ for a  classical network described by an acyclic directed graph $G'$, where $G$ is an undirected underlying graph of $G'$.
We showed that under the above assumptions, our protocol can multicast 
$1 \rightarrow 2$ ($1 \rightarrow 3$) asymmetric optimal UQCs of a $q^r$-dimensional state from the source node to the target nodes by consuming a small amount of entanglement that does not scale with $q$, which is shared among the target nodes. 

The extension of our protocol for $1 \rightarrow n$ asymmetric optimal UQCs for $n\le 4$ is not so straightforward. Hence, we leave this study as our future work.


\begin{thebibliography}{99}
\bibitem{HL08} Tracey Ho and Desmond S. Lun. {\it Network Coding: an introduction}. Cambridge University Press (2008)

\bibitem{Y08}Raymond W. Yeung, ``Information Theory and Network Coding'', Springer (2008)

\bibitem{ACLY00}R. Ahlswede, Ning Cai, Shuo-Yen Robert Li, Raymond W. Yeung, ''Network information flow'', IEEE Trans. on Inf. Theor. {\bf 46}, No.4 (2000)


\bibitem{HINRY07}
M.~Hayashi, K.~Iwama, H.~Nishimura, R.~Raymond, and S.~Yamashita, 
``{Quantum Network Coding},'' 
in {\em STACS 2007 SE - 52} (W.~Thomas and P.~Weil, eds.), vol.~4393 of 
{\em Lecture Notes in Computer Science}, pp.~610--621, Springer Berlin Heidelberg, 2007.

\bibitem{Hayashi07}
M.~Hayashi, 
``{Prior entanglement between senders enables perfect quantum network coding with modification},'' 
{\em Phys. Rev. A}, vol.~76, no.~4,~40301, 2007.

\bibitem{Shi06} Y.~Shi and E.~Soljanin. ``{On multicast in quantum networks}''
in {\em 40th Annual Conference on Information Sciences and Systems}, 
	page 871-876, 2006

\bibitem{Kobayashi09}
H.~Kobayashi, F.~{Le Gall}, H.~Nishimura, and M.~R\"{o}tteler, ``{General
  Scheme for Perfect Quantum Network Coding with Free Classical
  Communication},'' in {\em Automata, Languages and Programming SE - 52}
  (S.~Albers, A.~Marchetti-Spaccamela, Y.~Matias, S.~Nikoletseas, and
  W.~Thomas, eds.), 
  vol.~5555 of 
  {\em Lecture Notes in Computer Science},
  pp.~622--633, Springer Berlin Heidelberg, 2009.

\bibitem{Kobayashi10}
H.~Kobayashi, F.~{Le Gall}, H.~Nishimura, and M.~Rotteler, 
``{Perfect quantum network communication protocol based on classical network coding},'' 
in {\em Proceedings of 2010 IEEE International Symposium on Information Theory (ISIT)},
pp.~2686--2690, 2010.

\bibitem{Leung10}
D.~Leung, J.~Oppenheim, and A.~Winter, 
``{Quantum Network Communication; The Butterfly and Beyond},'' 
{\em IEEE Transactions on Information Theory}, vol.~56, no.~7,~3478--3490, 2010.


\bibitem{Kobayashi11}
H.~Kobayashi, F.~{Le Gall}, H.~Nishimura, and M.~Rotteler, ``{Constructing
  quantum network coding schemes from classical nonlinear protocols},'' 
in {\em Proceedings of 2011 IEEE International Symposium on Information Theory (ISIT)},
pp.~109--113, 2011.

\bibitem{OKM13}M.~Owari, G.~Kato, M.~Murao, ``{Multicast 
	quantum network coding on the butterfly network}'' Japan patent 
	JP2013-201654A (in Japanese)
\bibitem{KOM14}G.~Kato, M.~Owari, M.~Murao, ``{Multicast quantum 
	network coding}'' Japan patent JP2014-192875A (in Japanese)
\bibitem{KOM15}G.~Kato, M.~Owari, M.~Murao ``{Multicast quantum 
	netowk coding}'' Japan patent JP2015-220621A (in Japanese)
\bibitem{EKB16}Michael Epping, Hermann Kampermann, Dagmar Bru\ss,
``Quantum Router with Network Coding'', {\em New Journal of Physics}, vol.18, 103052 (2016)

\bibitem{OKH17a} M. Owari, G. Kato, and M. Hayashi, “Secure Quantum Network Coding
on Butterfly Network,” {\em Quantum Science and Technology}, vol. 3, 014001
(2017).

\bibitem{OKH17b} G. Kato, M. Owari, and M. Hayashi, “Single-Shot Secure Quantum
Network Coding for General Multiple Unicast Network with Free Public
Communication,” {\em In: Shikata J. (eds) 10th International Conference on
Information Theoretic Security (ICITS2017). Lecture Notes in Computer
Science}, vol 10681. Springer, pp. 166-187.

\bibitem{SH18b} Seunghoan Song, Masahito Hayashi
``Quantum Network Code for Multiple-Unicast Network with Quantum Invertible Linear Operations'',
{\em Proceedings of 13th Conference on the Theory of Quantum Computation, Communication and Cryptography (TQC 2018)} (2018)

\bibitem{SH18a} Seunghoan Song, Masahito Hayashi,
``Secure Quantum Network Code without Classical Communication''
arXiv:1801.03306 (2018)



\bibitem{SINV16} Takahiko Satoh, Kaori Ishizaki, Shota Nagayama, Rodney Van Meter,
``Analysis of Quantum Network Coding for Realistic Repeater Networks''
{\em Physical Review A} vol.93, 032302 (2016)



\bibitem{MSNV18} Takaaki Matsuo, Takahiko Satoh, Shota Nagayama, Rodney Van Meter,
``Analysis of Measurement-based Quantum Network Coding over Repeater Networks under Noisy Conditions'', {\em Physical Review A}, vol.97, 062328 (2018)

\bibitem{AM16}Seiseki Akibue, Mio Murao,
``Network coding for distributed quantum computation over cluster and butterfly networks''
{\em IEEE Transaction on Information Theory} vol.62, pp. 6620 - 6637 (2016)

\bibitem{TRLKM06}Danail Traskov  Niranjan Ratnakar ; Desmond S. Lun ; Ralf Koetter ; Muriel Medard
``Network Coding for Multiple Unicasts: An Approach based on Linear Optimization'',
{\em Proceedings of 2006 IEEE International Symposium on Information Theory (ISIT2006)}, pp. 1758-1762 (2006)

\bibitem{WZ82} William K. Wootters, Wojciech H. Zurek, Nature, 299 (1982)

\bibitem{BH96}Vladimir Buzek, Mark Hillery, Phys. Rev. A, 54, 1844 (1996)

\bibitem{SIG05} Valerio Scarani, Sofyan Iblisdir, and Nicolas Gison, ``Quantum cloning'', Rev. Mod. Phys. {\bf 77}, pp.1225-1256, (2005).

\bibitem{FWJYSZM14} Heng Fan, Yi-Nan Wang, Li Jing, Jie-Dong Yue, Han-Duo Shi, Yong-Liang Zhang, Liang-Zhu Mu, {\it Physics Reports} {\bf 544}, pp. 241-322, (2014).

\bibitem{NG98} Chi-Sheng Niu, Robert B. Griffiths, \emph{Phys. Rev. A}, {\bf 58}, 4377, (1998)

\bibitem{C98} Nicolas J. Cerf, \emph{Acta. Phys. Slov.}, {\bf 48}, 115, (1998)

\bibitem{C00} Nicolas J. Cerf, \emph{J. Mod. Opt.}, {\bf 47}, pp.187-209, (2000)

\bibitem{IACFFG05} S. Iblisdir, A. Ac\'in, N. J. Cerf, R. Filip, J. Fiur\'a\v{s}ek, and N. Gisin, \emph{Phys. Rev. A}, {\bf 72}, 042328 (2005).

\bibitem{Haya2}
M. Hayashi, \textit{Group Representation for Quantum Theory},
Springer (2017)


\bibitem{key2} Debbie Leung, Jonathan Oppenheim, Andreas Winter. ``Quantum network communication -- the butterfly and beyond'',  IEEE Transactions on Information Theory {\bf56}, pp.3478 - 3490, (2010) 

\bibitem{key3} Hirotada Kobayashi, Fran\c{c}ois Le Gall, Harumichi Nishimura, Martin R\"{o}tteler, ``General Scheme for Perfect Quantum Network Coding with Free Classical Communication'', In {\it ICALP 2009}, {\bf 5555} of {\it Lecture Note in Computer Science}, pp.622-633, (2009)

\bibitem{key4}  Hirotada Kobayashi, Fran\c{c}ois Le Gall, Harumichi Nishimura, Martin R\"{o}tteler, ``Constructing quantum network coding schemes from classical nonlinear protocols'',  Information Theory Proceedings (ISIT), 2011 IEEE International Symposium on, (2011)

\bibitem{key5} Hirotada Kobayashi, Fran\c{c}ois Le Gall, Harumichi Nishimura, Martin R\"{o}tteler, ``Perfect Quantum Network Communication Protocol Based on Classical Network Coding'', {\it Proceedings 2010 IEEE International Symposium on Information Theory (ISIT 2010)}, pp. 2686-2690, (2010).

\bibitem{key6} Go Kato, Masaki Owari, Mio Murao, ``Multicast quantum network coding'', Japan-Patent, Tokkai 2015-220621 (2015)	

\bibitem{key7} Valerio Scarani, Sofyan Iblisdir, and Nicolas Gison, ``Quantum cloning'', Rev. Mod. Phys. {\bf 77}, pp.1225-1256, (2005).

\bibitem{key8} Heng Fan, Yi-Nan Wang, Li Jing, Jie-Dong Yue, Han-Duo Shi, Yong-Liang Zhang, Liang-Zhu Mu, {\it Physics Reports} {\bf 544}, pp. 241-322, (2014).

\bibitem{key9} S. Iblisdir, A. Ac\'in, N. J. Cerf, R. Filip, J. Fiur\'a\v{s}ek, and N. Gisin, ``Multipartite asymmetric quantum cloning'' 
Phys. Rev. A {\bf 72}, 042328 (2005).

\bibitem{key10}Charles H. Bennett, Gilles Brassard, Claude Cr\'{e}peau, Richard jozsa, Asher Peres, and William K. Wootters, ``Teleporting an unknown quantum state via dual classical and Einstein-Podolsky-Rosen Channels'', Phys. Rev. Lett. {\bf 70}, (1993). 
\end{thebibliography}
\end{document}